\documentclass[12pt]{article}
\usepackage{setspace} 

\usepackage[english]{babel}

\usepackage{cmap}
\usepackage{indentfirst}
\usepackage{epsfig}

\usepackage{amsmath}
\usepackage{amssymb}
\usepackage{mathrsfs}
\usepackage{mathabx}
\usepackage{slashed}
\usepackage{nicefrac}
\usepackage{array}

\newcommand{\bracket}[2]{\big\langle \, #1 \big\vert \, #2\,\big\rangle}
\newcommand{\matrixel}[3]{\left <\, #1 \left |\, #2\,\right |\, #3\,\right >}

\begin{document}

\begin{center}
{\bf {\Large Decays of charged $B$--mesons into three charged leptons
    and a neutrino}}
\end{center}

\vspace{0.7cm}
\begin{center}
  {\large A.~Danilina$^{1,2,3}$, N.~Nikitin$^{1,2,3}$, K.~Toms$^{4}$} \\
   $^{1}$ Lomonosov Moscow State University, Department of Physics,
   Russia \\
   $^{2}$ Lomonosov Moscow State University, Skobeltsyn Institute of
          Nuclear Physics, Russia \\
   $^{3}$Institute for Theoretical and Experimental Physics, Russia \\
   $^{4}$Department of Physics and Astronomy, University of New
   Mexico, USA \\
\end{center}

\section*{Abstract}
In the framework of the Standard Model we present predictions for partial
widths, double and single differential distributions, and forward--backward
lepton asymmetries for four-leptonic decays
$B^- \to \mu^+\mu^- {\bar \nu}_e\, e^-$,  $B^- \to e^+ e^- {\bar
  \nu}_\mu\,\mu^-$, $B^- \to \mu^+ \mu^-{\bar \nu}_\mu\,\mu^-$, and
$B^- \to e^+  e^- {\bar \nu}_e\, e^-$. We consider the contributions
of virtual photon emission from the light and heavy quarks of the
$B^-$--meson, and we include bremsstrahlung of a virtual photon from
the charged lepton in the final state. We use the model of vector
meson dominance for calculation of virtual photon emission by the
light quark of the $B^-$--meson and take into account the isotopic
correction.


\section*{Introduction}

Four-leptonic decays of $B$--mesons allow a precise test of
Standard Model (SM) predictions in the higher orders of perturbation
theory. At the same time these decays may be background processes
to the helicity-suppressed ultra-rare decays
$B_{d,\, s} \to \mu^+ \mu^-$, which are under study at the
Large Hadron Collider (LHC)
\cite{CMS:2014xfa,Aaboud:2016ire,Aaij:2017vad}. These studies are
motivated by searches for Beyond the Standard Model physics.

Rare four-leptonic decays of $B$--mesons in the SM may be divided into
two groups. The decays of the first group are forbidden at the tree level
and occur through the higher order loop diagrams of perturbation
theory -- ``penguin'' and/or ``box''. In this way of the SM includes
flavor changing neutral currents (FCNC). An example of the first
group of decays is the process $B_s \to e^+ e^-\mu^+\mu^-$ and any other
four-leptonic decays of neutral $B$--mesons. In the second group, in
order to obtain the given multi-lepton final state, a number of
tree level weak and electromagnetic processes are involved. Examples
are the decay $B^- \to e^+ e^- {\bar \nu}_\mu\,\mu^-$ and
analogous processes involving charged $B$--mesons. Both groups are
studied at the LHC and potentially could be investigated at the Belle
II experiment. Currently only upper limits for branching ratios of the
decays $B_{d,\, s}\to \mu^+ \mu^- \mu^+ \mu^-$ and $B^- \to\, \mu^+
{\bar \nu}_{\mu}\, \mu^-  \mu^-$ are available \cite{Aaij:2013lla,
  Aaij:2016kfs,Aaij:2018pka}.

The experimental upper limits \cite{Aaij:2013lla, Aaij:2016kfs} for
the decays $B_{d,\, s}\to \mu^+ \mu^- \mu^+ \mu^-$ are an order of
magnitude higher than the corresponding theoretical predictions
\cite{Dincer:2003zq}  and estimates \cite{Danilina:2018uzr}.
The situation with the decay $B^- \to\, \mu^+ {\bar \nu}_{\mu}\, \mu^-
\mu^-$ is different. The experimental upper limit \cite{Aaij:2018pka},
\begin{eqnarray}
\label{LHCbB2mu_mu_mu_nu}
\textrm{Br} \left ( B^- \to\, \mu^+ {\bar \nu}_{\mu}\, \mu^-  \mu^- \right )\, <\, 0.16\,\times\, 10^{-7},
\end{eqnarray}
obtained with 95\% confidence level (CL) is almost an order of
magnitude lower than the theoretical predictions \cite{Danilina:2018uzr,Danilina:2018tsa}.
We present here to more detailed calculation of the branching ratios of
$B^- \to \mu^+\mu^- {\bar \nu}_e\, e^-$,  $B^- \to e^+ e^- {\bar
  \nu}_\mu\,\mu^-$, $B^- \to \mu^+ \mu^-{\bar \nu}_\mu\,\mu^-$ and
$B^- \to e^+  e^- {\bar \nu}_e\, e^-$, taking into account isotopic
effects. Also in the phase space of the decays, a correction to
non-zero lepton mass is considered. While this leads to better agreement
between theory and experiment, some discrepancy remain. Special
attention is given to the predictions of the behavior
of differential distributions, e.g. forward--backward lepton asymmetries.

This article is organized as follows. In the ``Introduction'' we give
a task description. In Section~\ref{sec:B2lllnuHeff} we write the
effective Hamiltonian and give definite the hadronic
form factors. In Section~\ref{sec:strukturaB2lllnu} the common
dependence of the decay amplitudes $B^-\,\to\, \ell^+ \ell^- {\bar
  \nu}_{\ell'}\,\ell'^-$ on di-lepton 4-momenta is
studied. Section~\ref{sec:B2lllnuTochnieFormuli} contains the exact
formulae for amplitudes of the decay $B^-\,\to\, \ell^+ \ell^- {\bar
  \nu}_{\ell'}\,\ell'^-$ for $\ell \ne \ell'$, and Section
~\ref{sec:B2lllnuTochnieFormuli2} provides analogous formulae for 
$\ell \equiv \ell'$. In Section~\ref{sec:B2lllnuNumericalResults} we
present numerical results for the decays of charged $B$--mesons into
three charged leptons and Á neutrino and discuss the precision of the
predictions. The ``Conclusion'' contains the main outcome of
the work. Some details of the four-leptonic decay kinematics are
given in Appendix~\ref{sec;kinemat4}.


\section{Effective Hamiltonian and hadronic matrix elements}
\label{sec:B2lllnuHeff}

In terms of fundamental quark and lepton fields, the Hamiltonian for
calculation of the amplitudes of four-lepton decays
$
B^-\,\to\, \ell^+\,\ell^-\,{\bar \nu}_{\ell'}\,\,\ell'^- 
$
has the form:
\begin{eqnarray}
\label{newHeff}
 {\cal H}_{\mathrm{eff}} (x)\, =\, {\cal H}_{W}(x)\, +\, {\cal H}_{\mathrm{em}}(x). 
\end{eqnarray}
The Hamiltonian of the transitions $b \to u W^- \to u \ell^- {\bar\nu}_{\ell}$
is written as:
$$
{\cal H}_{W}(x)\, =\, -\,\frac{G_F}{\sqrt{2}}\,V_{ub}\,
\Big (
\bar u(x)\,\gamma^{\mu} (1 - \gamma^5)\, b(x)
\Big )\,
\Big (
\bar\ell (x)\,\gamma_{\mu} (1 - \gamma^5) \nu_{\ell}(x)
\Big )\, +\, h.c.,
$$
where $u(x)$ and $b(x)$ are quark fields, $\ell(x)$ and
$\nu_{\ell}(x)$ are lepton fields, $G_F$ is the Fermi constant,
$V_{ub}$ is the corresponding matrix element of the
Cabibbo-Kobayashi-Maskawa (CKM) matrix, and the matrix $\gamma^5$ is
defined as $\gamma^5 = i \gamma^0 \gamma^1 \gamma^2 \gamma^3$. 

The Hamiltonian of the electromagnetic interaction has the form:
$$
{\cal H}_{\mathrm{em}}(x)\, =\, -\, e\,\sum\limits_f\, Q_f
\Big (
\bar f(x)\,\gamma^{\mu} f(x)
\Big )\,
A_{\mu}(x)\, =\, -\, j_{\mathrm{em}}^{\mu}(x)\,A_{\mu}(x),
$$
where the unitary charge $e = |e|$ is normalized by $e^2 = 4 \pi
\alpha_{em}$; $\alpha_{em} \approx 1/137$, the fine structure constant,
$Q_f$ is the charge of the fermion of flavor $f$ in units of the
unitary charge, $f(x)$ is the fermionic field of flavor $f$, and $A_{\mu}
(x)$ is the four-potential of the electromagnetic field. 

We define the following non-zero hadronic matrix elements, which are
needed for the subsequent calculations:
\begin{eqnarray}
\label{hadr_matrix_ell-B2lllnu}
\matrixel{0}{\bar u\,\gamma^{\mu} \gamma^5 b}{B^-(M_1,\, p)} &=& i\, f_{B_u}\, p^{\mu},
\nonumber \\
\matrixel{0}{\bar q\,\gamma^{\mu}  Q}{V(M_V,\, k,\, \varepsilon)} &=& \varepsilon^{\mu}\, M_V\, f_V,
\nonumber \\
\matrixel{V(M_2,\, q,\,\varepsilon)}{\bar u\,\gamma_{\mu}  b}{B^-(M_1,\, p)} &=& \frac{2\, V(k^2)}{M_1 + M_2}\,
               \epsilon_{\mu \nu \alpha \beta}\, \varepsilon^{*\, \nu} p^{\alpha} q^{\beta},
\\
\matrixel{V(M_2,\, q,\,\varepsilon)}{\bar u\,\gamma_{\mu} \gamma^5  b}{B^-(M_1,\, p)} &=& 
i\,\varepsilon^{*\, \nu}\,
\left [
(M_1 + M_2) A_1 (k^2) g_{\mu \nu}\, -\,\frac{A_2 (k^2)}{M_1 + M_2}\, (p +q)_{\mu} p_{\nu}\, -
\right .
\nonumber\\
&-& 
\left .
\frac{2 M_2}{k^2}\,\Big( A_3(k^2) - A_0 (k^2)\Big )\, k_{\mu}\, p_{\nu}
\right ],
\nonumber\\
\matrixel{B^{*\, -}(M_{B^*},\, k,\, \varepsilon)}{\bar b\,\gamma^{\mu} b}{B^-(M_1,\, p)} &=&
\frac{2\, V_b (q^2)}{M_1 + M_{B^*}}\,
               \epsilon_{\mu \nu \alpha \beta}\, \varepsilon^{*\, \nu} p^{\alpha} k^{\beta},
\nonumber
\end{eqnarray}
where $M_1$ -- $B^-$ is the meson mass, $p^{\mu}$ is the its four-momentum,
$M_{B*}$ is the $B^{*\, -}$--meson mass, $M_2$ -- mass of the light
($\rho^0(770)$ or $\omega (782)$) mesons,  $M_V=\{M_2,\, M_{B^*}\}$ are
masses of the intermediate vector mesons, and $\varepsilon^{\mu}$ are
their polarizations. Four--vectors $p^{\mu}$, $q^{\mu}$, and $k^{\mu}$
satisfy the conservation law:
$
p^{\mu}\, =\, q^{\mu}\, +\, k^{\mu}
$.
The components of the fully antisymmetric tensor $\epsilon^{\mu \nu \alpha
  \beta}$ are fixed by the condition $\epsilon^{0123} = -
\epsilon_{0123} = -1$, and $g_{\mu \nu}$ is the metric tensor in Minkovsky
space with $\textrm{diag}\, g_{\mu \nu} = \left ( 1,\, -1,\, -1,\, -1\right )$.


\section{Generic structure of the amplitudes for the decays $B^-\,\to\,
  \ell^+\,\ell^-\, {\bar \nu}_{\ell'}\,\ell'^-$ with the zero lepton mass
  approximation}
\label{sec:strukturaB2lllnu}

\begin{figure}[bt]
\begin{center}
\includegraphics[width=12.00cm]{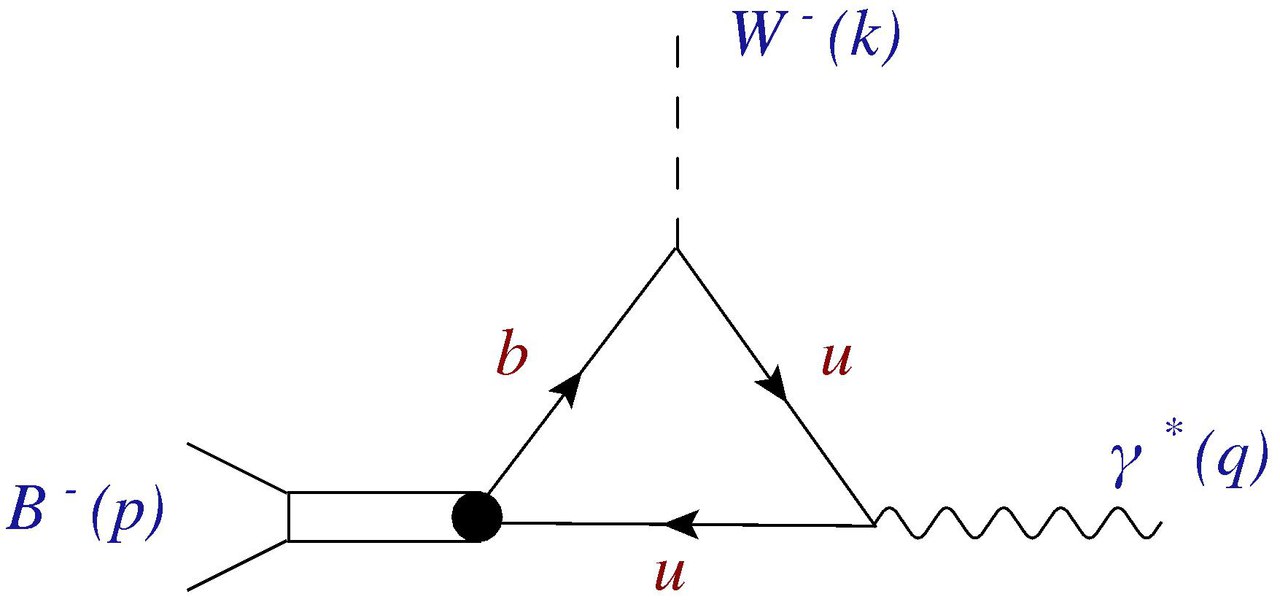}
\end{center}
\caption{\protect\label{fig:Fu} 
Emission of a virtual photon by the light quark of the $B^-$--meson.}
\end{figure}

\begin{figure}[bt]
\begin{center}
\includegraphics[width=12.00cm]{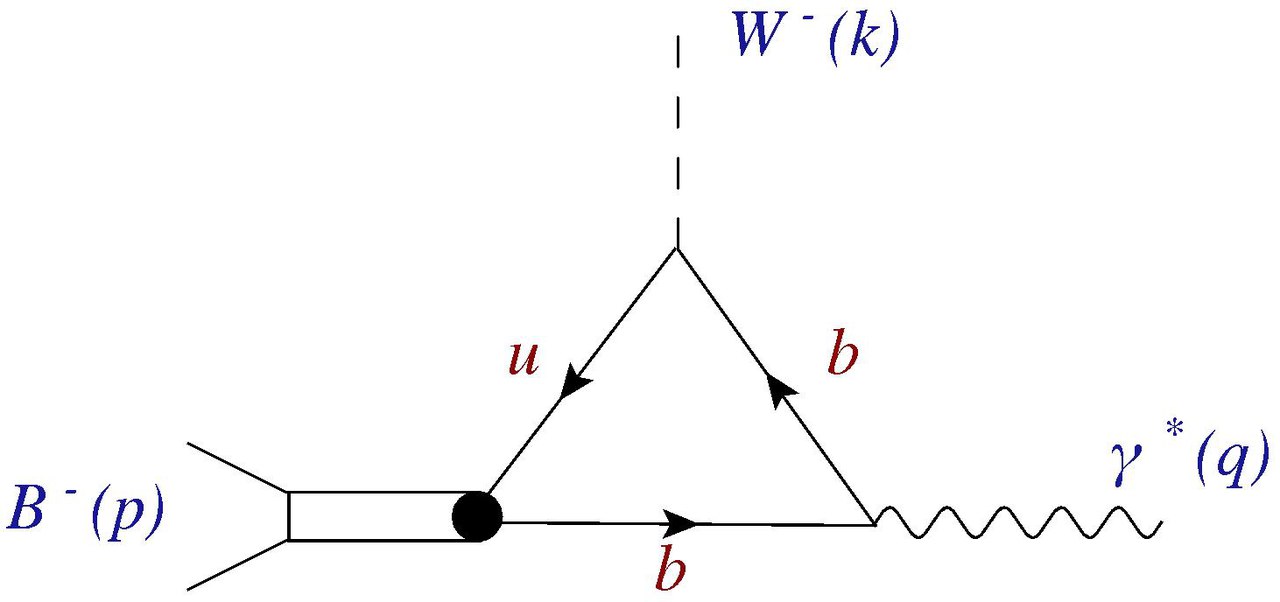}
\end{center}
\caption{\protect\label{fig:Fb} 
Emission of a virtual photon by the heavy quark of the $B^-$--meson.}
\end{figure}

There are three main types of diagrams needed for description of the decays 
$
B^-(p) \,\to\,\gamma^* (q)\, W^{-}(k)\,\to\,\ell^+(k_1)\,\ell^-(k_2)\, {\bar \nu}_{\ell'}(k_3)\,\ell'^-(k_4) 
$,
when the flavor of lepton $\ell$ is different from the flavor of
lepton $\ell'$. The first type arises in the situation when a virtual
photon is emitted by light a $u$--quark (see~Fig.~(\ref{fig:Fu})). The second type
corresponds to the emission of a virtual photon from a $b$--quark
(see~Fig.~(\ref{fig:Fb})). The third type is related to bremsstrahlung, when a
virtual photon is emitted by the lepton $\ell'^-$ in the final state
(see Fig.~(\ref{fig:FBR}) below). The four--momenta $q$ and $k$ are
defined in Appendix~\ref{sec;kinemat4}.

The structure of the amplitude, corresponding to diagrams on
Figs.~(\ref{fig:Fu}), (\ref{fig:Fb}) and (\ref{fig:FBR}) may be presented
as

\begin{eqnarray}
\label{Fu}
{\cal M}_{fi}\left (q^2,\, k^2\right )\, \sim\,\frac{1}{q^2}\, T_{\nu\mu}(q,\, k)\, j^{\nu}(k_2,\, k_1)\, J^{\mu} (k_4,\, k_3) ,
\end{eqnarray}
where
\begin{eqnarray}
T^{\nu\mu}(q,\, k) &=&
i\,\int\, d^4 x\, e^{i\, (qx)}
\matrixel{0}{j_{em}^\nu (x),\, \bar u(0)\, \gamma^{\mu}\, (1 - \gamma_5)\, b(0)}{B^- (M_1,\, p)}\, =\, \nonumber\\
&=&T^{(u)}_{\nu\mu}(q,\, k)\, +\,T^{(b)}_{\nu\mu}(q,\, k)\, +\,T^{(brem)}_{\nu\mu}(q,\, k).
\nonumber
\end{eqnarray}
The lepton currents are 
$$
j^{\nu} (k_2,\, k_1)\, =\,\Big (\overline{\ell}(k_2)\gamma^\nu\ell(-k_1) \Big )
\quad \textrm{and}\quad
J^{\mu} (k_4,\, k_3)\, =\,\Big (\overline{\ell\,'}(k_4)\gamma^\mu(1-\gamma^5)\nu_{\ell\,'}(-k_3) \Big ). 
$$
In the amplitude ${\cal M}_{fi}\left (q^2,\, k^2\right )$, the pole
$1/q^2$ of the photon propagator is evident. For calculations with
$q^2 \to 0$ it is necessary to take into account non-zero lepton
masses. This is done using the exact formula (\ref{dPhi1234}) for
four-particle phase space, and by introducing an effective cut for
some value $q^2_{\mathrm{min}}$. If $\ell \equiv \mu$ for $q^2_{\mathrm{min}}$ it makes
sense to choose the natural kinematical cut  $4m_{\mu}^2$. For the
case when $\ell \equiv e$ it is better to use the kinematical limits
of an experimental device, which are definitely higher than $4 m_e^2$.

Tensor $T_{\nu\mu}(q,\, k)$ satisfies the condition: $q^\nu\,
T_{\nu\mu}(q,\, k) = 0$. According to this condition, and taking
into account the result of Ref.\cite{Kozachuk:2017mdk}, tensor
$T_{\nu\mu}(q,\, k)$ has the form:
\begin{eqnarray}
\label{T-common}
T_{\nu\mu}(q,\, k)&=&\epsilon_{\nu\, \mu\, q\, k}\, \frac{e\, a\left (q^2,\, k^2\right )}{M_1}\, -\,
i\,\left ( g_{\nu\mu} - \frac{q_\nu\, q_\mu}{q^2}\right )\, e\, M_1\, b\left (q^2,\, k^2\right )\, - \\
&-&
i\, e\,\left ( k_\nu\, -\,\frac{(qk)}{q^2}\, q_\nu \right )\,\left
    (k_\mu\,\frac{2\, d\left (q^2,\, k^2\right )}{M_1}\, -\,q_\mu\,
    \frac{2\, c\left (q^2,\, k^2\right )}{M_1}\right )\, -\, i\,
    Q_{B_u}\, e\, f_{B_u}\frac{q_\nu k_\mu}{q^2},
\nonumber
\end{eqnarray}
where $Q_{B_u} = Q_b - Q_u = -1$, is the electric charge of the $B^-$--meson
in units of  $|e|$. The functions  $a\left (q^2,\, k^2\right )$,
$\ldots$ , $d\left (q^2,\, k^2\right )$ are dimensionless form factors
which depend on two variables, the squares of the transferred
four-momenta, $q^2$ and $k^2$. From (\ref{T-common}) it follows that 
$d(0,\, 0)\, =\,  Q_{B_u}\, f_{B_u}/M_1$.
 
Using the equations of motion, in the limit of massless leptons,
one can obtain the following generic structure for the ampliude ${\cal
  M}_{fi}$: 
\begin{eqnarray}
{\cal M}_{fi}\left (q^2,\, k^2\right ) &\sim& \frac{e}{q^2}\,
\left ( 
\epsilon_{\nu\, \mu\, q\, k}\, \frac{a\left (q^2,\, k^2\right )}{M_1}\, -\,
i\,g_{\nu\mu}\, M_1\, b\left (q^2,\, k^2\right )\, +\,
i\, k_\nu\, q_\mu\, \frac{2\, i\, c\left (q^2,\, k^2\right )}{M_1}
\right ) \nonumber\\
&&{}  \qquad j^{\nu}(k_2,\, k_1)\, J^{\mu} (k_4,\, k_3).
\nonumber
\end{eqnarray}
The exact calculation of the form factors $a\left (q^2,\, k^2\right )$,
$\ldots$, and $c\left (q^2,\, k^2\right )$ is quite complicated. In
the current work we will take into account only the leading singular
factors to the corresponding form factors. 

Let us start with a study of tensor $T^{(u)}_{\nu\mu}(q,\, k)$, which
describes the contribution of diagram from ~Fig.\ref{fig:Fu} to the tensor
$T_{\nu\mu}(q,\, k)$.
The main contribution to the structure of tensor $T^{(u)}_{\nu\mu}(q,\,
k)$ is given by the lightest intermediate vector resonances
that contain a $u \bar u$--pair. For such states tensor
$T^{(u)}_{\nu\mu}(q,\, k)$ has Breit--Wigner poles for variable
$q^2$. Taking into account only the contributions from $\rho^0
(770)$ and $\omega (782)$ mesons, we can write

\begin{eqnarray}
&& T^{(u)}_{\nu\mu}(q,\, k)\, \to 
\nonumber \\
&\to& \sum\limits_{i = \rho^0,  \omega}
\matrixel{0}{\bar u \gamma_\nu u}{V(M_{2i},  q, \epsilon)}\,
\frac{e}{M_{2i}^2 -  q^2 - i M_{2i} \Gamma_{2i}}
\matrixel{V(M_{2i}, q, \epsilon)}{\bar u \gamma_\mu (1 - \gamma^5) b}{B^-(M_1,\, p)},
\nonumber
\end{eqnarray}
where $M_{2i}$ and $\Gamma_{2i}$ are the masses and widths, respectively, of the
intermediate vector resonances. 

For the zero leptonic mass approximation, the range of values of
the variable $k^2$ is $0 \le k^2 \le M_1^2$. The closest pole in $k^2$
is related to the appearance of the intermediate vector state $B^{*\,
  -}$. As $M_{B^{*\, -}} > M_1$, this pole lies outside of the
kinematically allowed range of the decay $B^-\,\to\, \ell^+\,\ell^-\,
{\bar \nu}_{\ell'}\,\ell'^-$. The existence of the pole at the mass of
the $B^{*\, -}$--meson is taken into account when choosing the pole
parametrisation of the form factors of the transitions $B \to \rho$ and
$B \to \omega$ \cite{Melikhov:2000yu}. For non-zero leptonic masses,
$m^2_{\ell'} \le k^2 \le (M_1 - 2 m_\ell)^2$. Hence all the remarks
above on the poles of tensor $T^{(u)}_{\nu\mu}$ for variable $k^2$ are
still valid.

As the contribution from $\rho^0$ and $\omega$ resonances is
dominant, it is possible to use the following estimate for the
branching ratio of $B^-\,\to\, \mu^+\mu^- {\bar\nu}_{e}\, e^-$:
\begin{eqnarray}
\label{Bu2lllnu_Estimation}
&&\textrm{Br}^{(u)}(B^- \rightarrow \mu^+\mu^- {\bar\nu}_{e}\, e^-)\,\approx\\
&\approx&\left | \sqrt{\textrm{Br}(B^-\rightarrow \rho^0 e^-{\bar\nu}_{e})\,\textrm{Br}(\rho^0 \rightarrow \mu^+ \mu^-)}\, +\,
                           \sqrt{\textrm{Br}(B^-\rightarrow \omega e^-{\bar\nu}_{e})\,\textrm{Br}(\omega \rightarrow \mu^+ \mu^-)}\,\right  |^2
\approx\, \nonumber \\
&\approx& 0.3\,\times\,10^{-7},\nonumber
\end{eqnarray}
where the necessary experimental values for the branching ratios are taken
from \cite{Tanabashi:2018oca}. The estimate
(\ref{Bu2lllnu_Estimation}) does not take into account the fact that
the $\rho^0(770)$--meson is a wide resonance, i.e., in the case of the 
$\rho^0(770)$--meson, the naive factorization approximation should lead to a 
lower branching ratios. Also in estimate (\ref{Bu2lllnu_Estimation}), the
photon pole, which should also lead to
lower results, is not taken into account. Does estimate (\ref{Bu2lllnu_Estimation})
contradict the experimental upper limit in (\ref{LHCbB2mu_mu_mu_nu})?
We do not think so, because we attribute to the factor of two $2$
accuracy. But the estimate of (\ref{Bu2lllnu_Estimation}) does
point to the possibility that the minimum of the possible
theoretical predictions may be above the experimental limit \cite{Aaij:2018pka}.

Now consider tensor $T^{(b)}_{\nu\mu}(q,\, k)$, which is related to
diagram from Fig.~(\ref{fig:Fb}). In the limit of massless leptons
there are no poles for the variable $q^2$ in the kinematically allowed
range $0 \le q^2 \le  M^2_1$ for the tensor $T^{(b)}_{\nu\mu}(q,\,
k)$. The closest pole outside the allowed range corresponds to the $b \bar
b$ quark composition. This is the $\Upsilon (1S)$ meson, whose mass is 
almost two times higher than the mass of the $B^-$--meson. The dominant
contribution to emission of a virtual photon by a heavy quark is
described using the process 
$B^- \to B^{*\, -} \gamma^*$. In this case 
\begin{eqnarray}
 && T^{(b)}_{\nu\mu}(q,\, k)\, \to 
\nonumber \\
&\to& 
\matrixel{0}{\bar u \gamma_\mu (1 - \gamma^5) b}{B^{*\, -}(M_{B^*},  k, \epsilon)}\,
\frac{e}{M_{B^*}^2 -  k^2 }
\matrixel{B^{*\, -}(M_{B^*},  k, \epsilon)}{\bar b \gamma_\nu b}{B^-(M_1,\, p)}.
\nonumber
\end{eqnarray}
Note that the imaginary addition $- i M_{B^*} \Gamma_{B^*}$ does not
exist in the propagator, as $k^2 <  M_{B^*}^2$, i.e., the pole of the 
$B^*$--meson is not reached. The contribution from the $\Upsilon(1S)$ is
taken into account effectively when introducing pole parametrization
for the form factor $V_b(q^2)$.  For the variable $k^2$ in the
kinematically allowed range, the tensor $T^{(b)}_{\nu\mu}(q,\, k)$ does
not have any other poles.

Numerically the contribution of the process on the Fig.~(\ref{fig:Fb}) to the
branching ratio associated with the four-leptonic decay is suppressed comparing to the
contribution of the process on the Fig.~(\ref{fig:Fu}) by factor $\left (\Lambda /
  m_b\right )^2$, where $m_b \sim 5$ GeV, the mass of $b$--quark, and
the parameter $\Lambda \approx 300--500$ MeV. This follows from the
exact equations for the form factors of the rare leptonic radiative
decays of $B$--mesons \cite{Kruger:2002gf,Melikhov:2004mk}. Due to the interference between diagrams~(\ref{fig:Fu})
and~(\ref{fig:Fb}) near the photonic pole it is necessary, however to take into
account the contribution of the diagram (\ref{fig:Fb}) to the full branching ratio.

\begin{figure}[bt]
\begin{center}
\includegraphics[width=12.00cm]{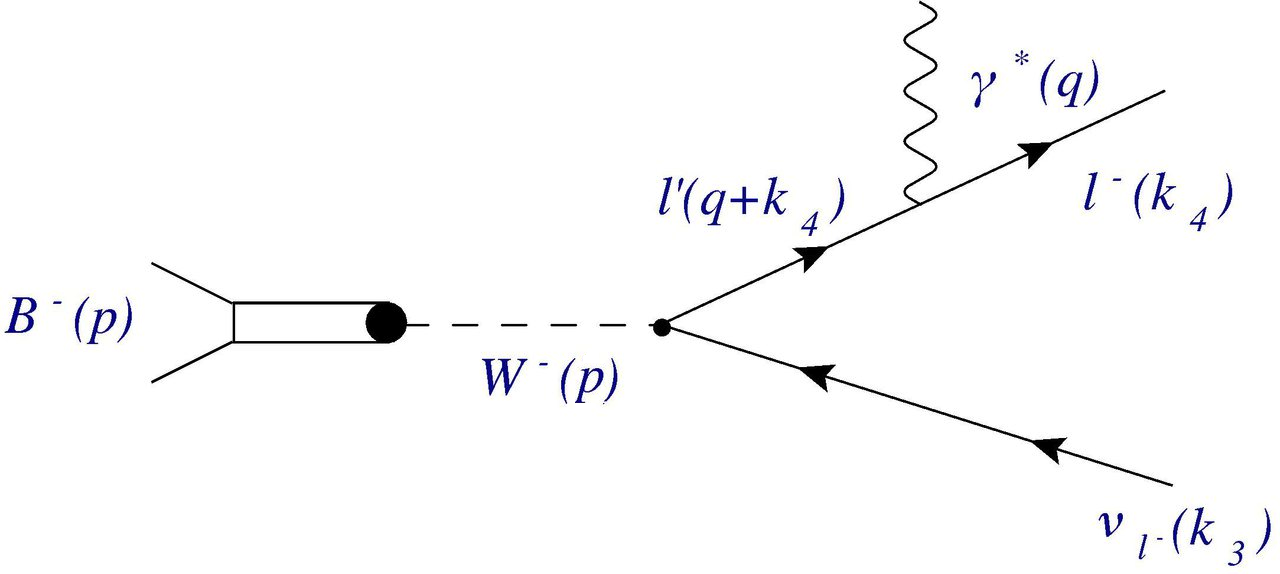}
\end{center}
\caption{\protect\label{fig:FBR} 
Bremsstrahlung of the virtual photon.}
\end{figure}

The bremsstrahlung contribution is described by diagram on the
Fig.~(\ref{fig:FBR}).
The bremsstrahlung amplitude has a single pole by $q^2$ from the
photon propagator. Hence the tensor $T^{(brem)}_{\nu\mu}(q,\, k)$ does
not have poles by $q^2$ and $k^2$. It is important to take into
account the bremsstrahlung contribution near the pole by $q^2$, where
the zero-mass approximation may not be fully correct. This
contribution should be calculated for non-zero lepton masses.


\section{Formulae for the decay $B^- \to\, \ell^+ \ell^-\, {\bar \nu}_{\ell'}\,\ell'^-$}
\label{sec:B2lllnuTochnieFormuli}

Consider the decays $B^- \to\, \mu^+ \mu^- {\bar \nu}_{e}\, e^-$ and
$B^- \to\, e^+ e^- {\bar \nu}_{\mu}\,\mu^-$, for the case when the
lepton flavors in the final state are different. Generally these decays
may be written as $B^- \to\, \ell^+ \ell^-\, {\bar
  \nu}_{\ell'}\,\ell'^-$ for $\ell \ne \ell'$. 

\begin{figure}[t!]
\begin{center}
\includegraphics[width=12.00cm]{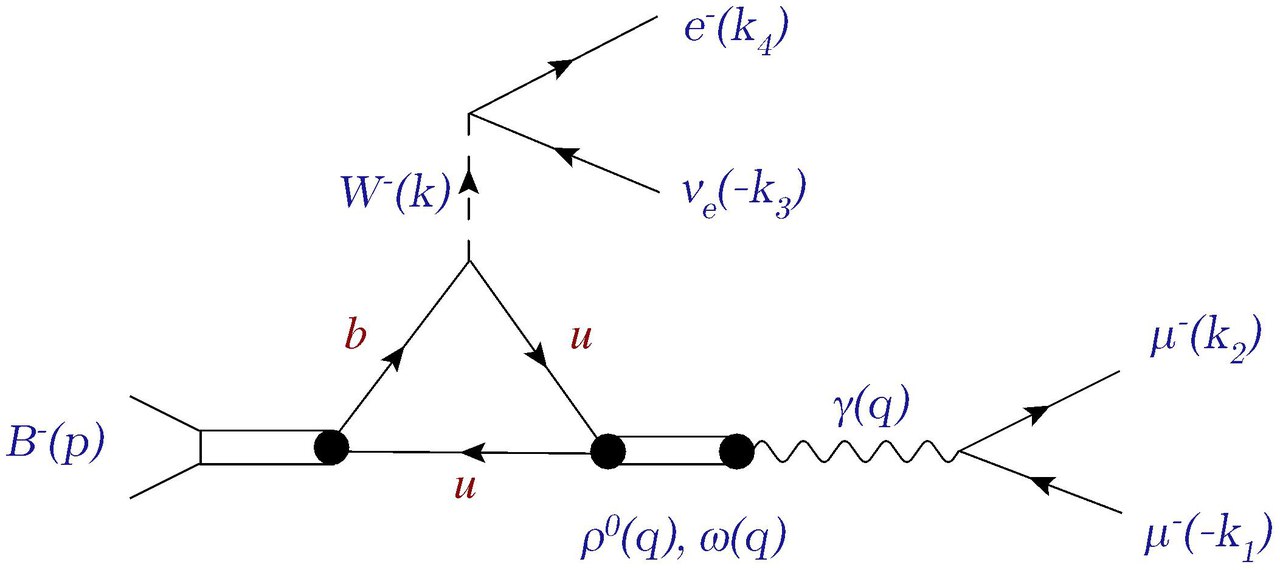} 
\end{center}
\caption{\label{fig:VMD} Diagram for calculation of ${\cal
    M}_{fi}^{(u)}$  (see  the equation (\ref{M_VMD})) using the decay
  $B^-\rightarrow \mu^+\mu^- {\bar\nu}_{e}\, e^-$ as an example. The
  emission of the virtual photon by a light quark is described by the 
  Vector Meson Dominance model.}
\end{figure}

The contribution to the full decay amplitude 
$
B^-(p) \,\to\, \ell^+(k_1)\,\ell^-(k_2)\, {\bar \nu}_{\ell'}(k_3)\,\ell'^-(k_4) 
$ 
from Fig.~(\ref{fig:Fu}) may be calculated using the Vector Meson
Dominance model (VMD) (see Fig.~(\ref{fig:VMD})). Assuming
$m_{\ell} = m_{\ell'} = 0$ and using the effective Hamiltonian (\ref{newHeff}), one finds, that for VMD the contribution
from process (\ref{fig:Fu}) is described by diagram
(\ref{fig:VMD}),  and the corresponding amplitude may be written as: 
\begin{eqnarray}
\label{M_VMD}
{\cal M}_{fi}^{(u)}&=& \frac{{\cal A}}{q^2}\,
\left [
\sum\limits_{i= \rho^0,\,\omega}
\,\frac{I_i\, M_{2i}\, f_{V_i}}{q^2-M_{2i}^2+i\Gamma_{2i} M_{2i}}{\cal F}_{\mu\nu}^{(i)}(k^2) 
\right ]\, j^{\nu} (k_2,\, k_1)\, J^{\mu} (k_4,\, k_3),
\end{eqnarray} 
where, using the motion equations,
$$
{\cal F}_{\mu\nu}^{(i)}(k^2)\, =\,
\frac{2\, V^{(i)}(k^2)}{M_1 + M_{2i}}\,
                 \epsilon_{\mu \nu k q}\, -\,
                i \, (M_1 + M_{2i}) A_1^{(i)} (k^2) g_{\mu \nu}\, +\, 2 i\,\frac{A_2^{(i)} (k^2)}{M_1 + M_{2i}}\, q_{\mu} k_{\nu}.
$$
For the calculation of the resonances sum in the (\ref{M_VMD}), only
the contributions from the lightest
$\rho^0$ and $\omega$ mesons, containing $u\bar
u$--pairs, are taken into account. Because the $\rho^0$ and $\omega$ mesons
are linear combinations of $u \bar u$ and $d \bar d$--pairs, in
order to extract the contributions only $u \bar u$--pair alone, an
isotopic coefficients $I_i$ are introduced. By definition 

$$I_{\rho^0} = \bracket{\rho^0}{\bar u\, u} = 1/\sqrt{2}$$ and  $$I_{\omega} = \bracket{\omega}{\bar u\, u} = 1/\sqrt{2}.$$

\begin{figure}[t!]
\begin{center}
\includegraphics[width=12.00cm]{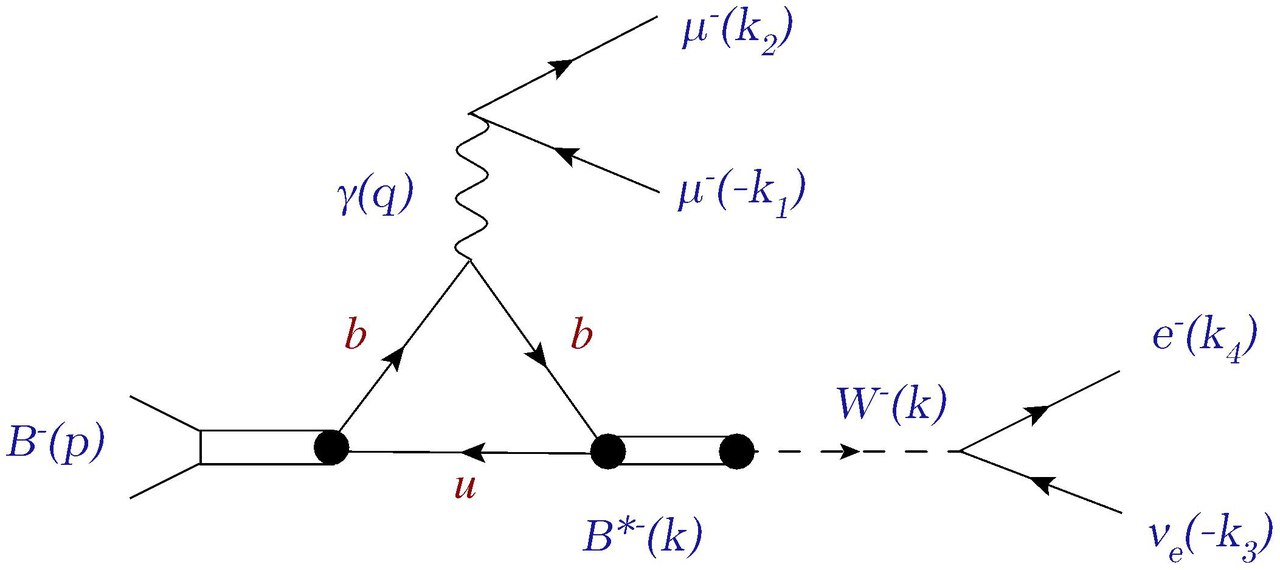}
\end{center}
\caption{\label{fig:BstarGamma} Diargam for the calculation of ${\cal
    M}_{fi}^{(b)}$  (see the equation (\ref{M_b})) using the decay
  $B^-\rightarrow \mu^+\mu^- {\bar\nu}_{e}\, e^-$ as an example.}
\end{figure}

The contribution from process (\ref{fig:Fb}) is given by
diagram on Fig.~(\ref{fig:BstarGamma}), which is the cross--channel of the decay
$B^* \to B \gamma^*$ of a heavy vector meson into a heavy pseudoscalar
meson and a virtual photon, and is represented by
\begin{eqnarray}
\label{M_b}
{\cal M}_{fi}^{(b)} &=&
\frac{2}{3}\,\frac{{\cal A}}{q^2}\,
\,\frac{M_{B^*} f_{B^*}}{k^2-M_{B^*}^2}\,
\frac{V_b (q^2)}{M_1+M_{B^*}}\,\epsilon_{\mu \nu k q}\,\, j^{\nu} (k_2,\, k_1)\, J^{\mu} (k_4,\, k_3).
\end{eqnarray}
There is no imaginary correction in the propagator, as $k^2 <  M_{B^*}^2$.

\begin{figure}[t!]
\begin{center}
\includegraphics[width=12.00cm]{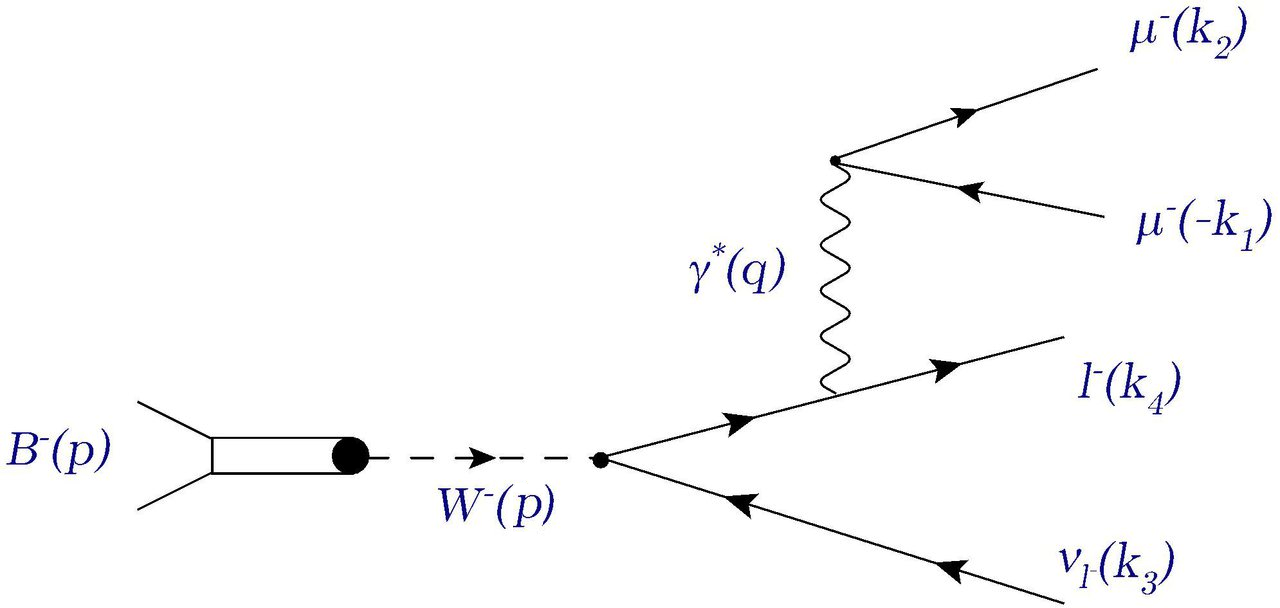} 
\end{center}
\caption{\label{fig:BremContrib} Diagram for the calculation of the
  amplitude of the bremsstrahlung ${\cal M}_{fi}^{(brem)}$  (see the equation (\ref{M_brem})) for the decay $B^-\rightarrow \mu^+\mu^-{\bar\nu}_{e}\, e^-$.}
\end{figure}

Finally, the contribution of the bremsstrahlung process (\ref{fig:FBR})
of the virtual photon is described by
diagram from Fig.~(\ref{fig:BremContrib}). In the case when $m_\ell \ne 0$ and
$m_{\ell'} \ne 0$, for the amplitude of the bremsstrahlung is:
\begin{eqnarray}
\label{M_brem-ml}
{\cal M}_{fi}^{(brem)}&=&\frac{{\cal A}}{q^2}\,i\, f_{B_u}\, g_{\mu\nu}\, j^{\nu} (k_2,\, k_1)\, {\tilde J}^{\mu} (k_4,\, k_3),
\nonumber
\end{eqnarray}
where
$$
{\tilde J}^{\mu} (k_4,\, k_3)\, =\, J^{\mu} (k_4,\, k_3)\, +\,\frac{m_{\,\ell'}}{(p - k_3)^2 - m_{\,\ell'}^2}\,
\,\Big (\overline{\ell\,'}(k_4)\,\gamma^\mu \left (\hat p  +  m_{\,\ell'} \right ) (1-\gamma^5)\,\nu_{\ell\,'}(-k_3) \Big ).
$$
As $(2 m_\ell + m_{\ell'})^2 \le (p - k_3)^2 \le M^2_1$, the second
summand does not contain any poles in the whole kinematically allowed
range. The second summand may be compatible with the first one only in
the range where $(p - k_3)^2 \sim (2 m_\ell + m_{\ell'})^2$. But this
range is suppressed by the phase space
(\ref{dPhi1234}) integration. For this the reason we assume that the
bremsstrahlung amplitude may be written as
\begin{eqnarray}
\label{M_brem}
{\cal M}_{fi}^{(brem)}&=&\frac{{\cal A}}{q^2}\,i\, f_{B_u}\, g_{\mu\nu}\, j^{\nu} (k_2,\, k_1)\, J^{\mu} (k_4,\, k_3).
\end{eqnarray} 
In formulas (\ref{M_VMD}), (\ref{M_b}) and (\ref{M_brem}) we denote
$
\displaystyle{\cal A}\, =\,\frac{G_F}{\sqrt{2}}\, 4\,\pi\,\alpha_{em}\, V_{ub}
$.

The full decay amplitude 
$
B^-(p) \,\to\, \ell^+(k_1)\,\ell^-(k_2)\, {\bar \nu}_{\ell'}(k_3)\,\ell'^-(k_4) 
$ 
may be written as
\begin{eqnarray}
\label{M1234}
{\cal M}_{fi}^{(1234)} &=& {\cal M}_{fi}^{(u)}\, +\, {\cal M}_{fi}^{(b)}\, +\, {\cal M}_{fi}^{(brem)}\, = \\
&=& \frac{{\cal A}}{q^2}\,
\left [
\frac{a(q^2, k^2)}{M_1}\,\varepsilon_{\mu\,\nu\, k\, q}\, -\, i M_1 b(q^2, k^2)\, g_{\mu \nu}\, +\, 2i\,\frac{c(q^2, k^2)}{M_1}\, q_{\mu} k_{\nu}
\right ]\,  j^{\nu} (k_2,\, k_1)\, J^{\mu} (k_4,\, k_3). \nonumber
\end{eqnarray}
The benefit of choosing the notation ${\cal M}_{fi}^{(1234)}$ for the
full amplitude will be apparent while considering decays with
identical charged fermions in the final state (see
Section~\ref{sec:B2lllnuTochnieFormuli2}). Dimensionless functions
$a(q^2, k^2) \equiv a (x_{12},\, x_{34})$, $b(q^2, k^2)\equiv b
(x_{12},\, x_{34})$, and $c(q^2, k^2) \equiv c (x_{12},\, x_{34})$ are
defined as:
\begin{eqnarray}
\label{abc_x12x34}
a(x_{12},\, x_{34}) &=& 
\frac{1}{3}\,\frac{\hat M_{B^*} \hat f_{B^*}}{x_{34}-\hat M_{B^*}^2}\,\frac{2\, V_b \left (M_1^2\,  x_{12} \right )}{1+ \hat M_{B^*}}\, + 
\nonumber \\
&+& \sum\limits_{i= \rho^0,\,\omega}
\frac{I_i\,\hat M_{2i}\,\hat f_{V_i}}{x_{12} - \hat M_{2i}^2+i\hat \Gamma_{2i} \hat M_{2i}}\,\frac{2\, V^{(i)}(M_1^2\, x_{34})}{1 + \hat M_{2i}};
\nonumber \\
b(x_{12},\, x_{34}) &=& - \hat f_{B_u}\, +\,
\sum\limits_{i= \rho^0,\,\omega}
\frac{I_i\,\hat M_{2i}\,\hat f_{V_i}}{x_{12}- \hat M_{2i}^2+i\hat \Gamma_{2i} \hat M_{2i}}\,(1 + \hat M_{2i})\, A_1^{(i)}(M_1^2\, x_{34});
\\
c(x_{12},\, x_{34}) &=&
\sum\limits_{i= \rho^0,\,\omega}
\frac{I_i\,\hat M_{2i}\,\hat f_{V_i}}{x_{12}- \hat M_{2i}^2+i\hat \Gamma_{2i} \hat M_{2i}}\,\frac{A_2^{(i)}(M_1^2\, x_{34})}{1 + \hat M_{2i}},
\nonumber
\end{eqnarray} 
where the dimensionless variables $x_{12}= q^2/M_1^2$, and $x_{34} =
k^2/M_1^2$ are defined in Appendix~\ref{sec;kinemat4}, and the 
dimensionless constants are defined as $\hat f_{B_u} = f_{B_u}/M_1$,
$\hat f_{B^*} = f_{B^*}/M_1$,  $\hat f_{V_i} = f_{V_i}/M_1$, $\hat
M_{2i} = M_{2i}/M_1$, $\hat M_{B^*} = M_{B^*}/M_1$ and $\hat \Gamma_{2i}
= \Gamma_{2i}/M_1$. Form factors $V_b(q^2)$, $V^{(i)} (k^2)$,
$A_1^{(i)} (k^2)$, and $A_2^{(i)} (k^2)$ are also dimensionless
functions.

The differential branching ratio of the decay 
$
B^- \,\to\, \ell^+\,\ell^-\, {\bar \nu}_{\ell'}\,\ell'^-
$
is calculated as
\begin{eqnarray}
\label{dBr1234common}
d \textrm{Br} \left (B^-\,\to\, \ell^+ \ell^- {\bar \nu}_{\ell'} \ell'^- \right ) \, =\, \tau_{B^-}\,
\frac{\sum\limits_{s_1,\, s_2,\, s_3,\, s_4}\left | {\cal M}_{fi}^{(1234)} \right |^2}{2 M_1}\, d\Phi_4^{(1234)}
\end{eqnarray}
where $\tau_{B^-}$ is the lifetime of the $B^-$--meson, four-particle
phase space $d\Phi_4^{(1234)}$ is defined by Equation
(\ref{dPhi1234}), and the summation is performed over the spins of the
final fermions. In formula (\ref{dBr1234common}) the integration over
the angular variables $y_{12}$, $y_{34}$, and $\varphi$ may be
performed analytically. After the integration,
\begin{eqnarray}
\label{dBr1234/dx12dx34}
&& \frac{d^2\,\textrm{Br} \left (B^-\,\to\, \ell^+ \ell^- {\bar \nu}_{\ell'} \ell'^- \right )}{d x_{12}\, d x_{34}}\, =\, 
\tau_{B^-}\,\frac{G_F^2\, M_1^5\, \alpha_{em}^2\, |V_{ub}|^2}{2^6\, 3^2\, \pi^3}\,
\sqrt{1\, -\,\frac{4 {\hat m}_\ell^2}{x_{12}}}\,\,\left ( 1\, -\,\frac{{\hat m}_{\ell'}^2}{x_{34}}\right )\\
&&\frac{\lambda^{1/2}(1,\, x_{12},\, x_{34})}{x_{12}^2}\,\Big [
2\, x_{12} x_{34}\,\lambda (1,\, x_{12},\, x_{34})\,\Big | a(x_{12}, x_{34}) \Big |^2\, +\nonumber \\
&+& 
\big (\lambda (1,\, x_{12},\, x_{34}) \, +\, 12\, x_{12}\, x_{34} \big )\,\Big | b(x_{12}, x_{34}) \Big  |^2\, +\,
\lambda^2 (1,\, x_{12},\, x_{34}) \,\Big | c (x_{12}, x_{34}) \Big |^2\, + \nonumber \\
&+& 2\,\lambda (1,\, x_{12},\, x_{34}) \, (x_{12} + x_{34} - 1)\,\textrm{Re}\, \Big ( b(x_{12}, x_{34})\, c^*(x_{12}, x_{34})\Big )
\Big ], \nonumber
\end{eqnarray}
which depends only on the dimensionless variables $x_{12}$ and
$x_{34}$. Full integration by these variables may be performed only
numerically. 

Because in the decay of the $B^-$--meson, all the leptons in the final state
are different, it makes sense to define two forward--backward leptonic
asymmetries $A^{(B^-)}_{FB}(x_{12})$ and $A^{(B^-)}_{FB}(x_{34})$ as
\begin{eqnarray}
\label{Afbq2-def}
A^{(B^-)}_{FB}(x_{12})\, =\,
\frac{\int\limits_0^1
d \cos\,\tilde\theta_{12}\,\frac{\displaystyle d^2\,\Gamma \left (B^-\,\to\, \ell^+ \ell^- {\bar \nu}_{\ell'} \ell'^- \right )}{\displaystyle d x_{12}\, d \cos\,\tilde\theta_{12}}\,-\,
\int\limits_{-1}^0
d \cos\,\tilde\theta_{12}\,\frac{\displaystyle d^2\,\Gamma \left (B^-\,\to\, \ell^+ \ell^- {\bar \nu}_{\ell'} \ell'^- \right )}{\displaystyle d x_{12}\, d \cos\,\tilde\theta_{12}}
}
{ \frac{\displaystyle d\,\Gamma \left (B^-\,\to\, \ell^+ \ell^- {\bar \nu}_{\ell'} \ell'^- \right )}{\displaystyle d x_{12}}}
\end{eqnarray} 
and
\begin{eqnarray}
\label{Afbk2-def}
A^{(B^-)}_{FB}(x_{34})\, =\,
\frac{\int\limits_0^1
d \cos\,\tilde\theta_{34}\,\frac{\displaystyle d^2\,\Gamma \left (B^-\,\to\, \ell^+ \ell^- {\bar \nu}_{\ell'} \ell'^- \right )}{\displaystyle d x_{34}\, d \cos\,\tilde\theta_{34}}\,-\,
\int\limits_{-1}^0
d \cos\,\tilde\theta_{34}\,\frac{\displaystyle d^2\,\Gamma \left (B^-\,\to\, \ell^+ \ell^- {\bar \nu}_{\ell'} \ell'^- \right )}{\displaystyle d x_{34}\, d \cos\,\tilde\theta_{34}}
}
{ \frac{\displaystyle d\,\Gamma \left (B^-\,\to\, \ell^+ \ell^- {\bar \nu}_{\ell'} \ell'^- \right )}{\displaystyle d x_{34}}},
\end{eqnarray} 
where $\tilde\theta_{12}$ is the angle between the
propagation directions of the $\ell^-$ and $B^-$ in the rest frame of the 
$\ell^+\ell^-$--pair, and $\tilde\theta_{34}$ is the angle between
the propagation directions of $\ell'^{\,-}$ and $B^-$ in the rest
frame of the $\ell'^{\, -}\bar\nu_{\ell'}$ pair. It is obvious that
$\tilde\theta_{12} = \pi - \theta_{12}$ and $\tilde\theta_{34} = \pi -
\theta_{34}$. Equations (\ref{Afbq2-def}) and
(\ref{Afbk2-def}) are chosen such that they correspond to the
notions of Ref. \cite{Kozachuk:2017mdk}.


\section{Exact formulae for the decay $B^- \to\, \ell^+  {\bar \nu}_{\ell}\,\ell^- \ell^-$}
\label{sec:B2lllnuTochnieFormuli2}

In practice, the muonic tracks are registered with a much higher
efficiency at almost all contemporary experiments. That is why from
the experimental point of view the decay $B^- \to\, \mu^+ {\bar
  \nu}_{\mu}\,\mu^-\mu^-$ is of the most interest. In this decay the
final state contains two identical muons of negative charge. Hence
the Fermi antisymmetry should be taken into account. 

Consider the full amplitude of the decay 
$
B^-(p) \,\to\, \ell^+(k_1)\, {\bar \nu}_{\ell}(k_3)\,\ell^-(k_2)\,\ell^-(k_4) 
$.
In the approximation of zero leptonic masses, the calculation below is
applicable to the decay $B^- \to\, \mu^+  {\bar
  \nu}_{\mu}\,\mu^-\mu^-$ as well as to the decay $B^- \to\, e^+
{\bar \nu}_{e}\, e^- e^-$. The full amplitude of the decay may be
written as 
\begin{eqnarray}
\label{M1234-M1432}
{\cal M}_{fi}^{(tot)}\, =\, {\cal M}_{fi}^{(1234)}\, -\,{\cal M}_{fi}^{(1432)},
\end{eqnarray}
where the amplitude ${\cal M}_{fi}^{(1234)}$ is set by equation
(\ref{M1234}), and the amplitude ${\cal M}_{fi}^{(1432)}$ can be
obtained from ${\cal M}_{fi}^{(1234)}$ by exchanging $k_2
\leftrightarrow k_4$. This leads to the necessity of replacing
$q_{\mu} \to \tilde q_{\mu}$, $k_{\mu} \to \tilde k_{\mu}$,  $x_{12}
\to x_{14}$, and $x_{34} \to x_{23}$ (see Appendix \ref{sec;kinemat4})
in the calculation of ${\cal M}_{fi}^{(1432)}$.

The differential branching ratio of the decay is given by 
\begin{eqnarray}
\label{dBr_common}
d \textrm{Br} \left (B^- \right . &\to& \left . \ell^+ {\bar \nu}_{\ell}\, \ell^-  \ell^- \right ) \, =\,\frac{1}{2}\,
\left [
\tau_{B^-}\,\frac{\sum\limits_{s_1,\, s_2,\, s_3,\, s_4}\left | {\cal M}_{fi}^{(1234)} \right |^2}{2 M_1}\, d\Phi_4^{(1234)}\, + 
\right .
\nonumber \\
&+& \tau_{B^-}\,\frac{\sum\limits_{s_1,\, s_2,\, s_3,\, s_4}\left | {\cal M}_{fi}^{(1432)} \right |^2}{2 M_1}\, d\Phi_4^{(1432)}\, -
\\
&-&
\left .
\tau_{B^-}\,\frac{\sum\limits_{s_1,\, s_2,\, s_3,\, s_4}\,\left (
{\cal M}_{fi}^{(1234)\,\dagger}\,{\cal M}_{fi}^{(1432)}\, +\, {\cal M}_{fi}^{(1432)\,\dagger}\,{\cal M}_{fi}^{(1234)}
\right )}{2 M_1}\,\, d\Phi_4^{(1234)}
\right ],
\nonumber
\end{eqnarray}
where $d\Phi_4^{(1234)}$ and $d\Phi_4^{(1432)}$ are set by
equations (\ref{dPhi1234}) and (\ref{dPhi1432}).The common factor of
$1/2$ is due to by Fermi antisymmetry.  

The first and the second summands in (\ref{dBr_common}) are equal. Hence for the branching ratio, it
is possible to write 
\begin{eqnarray}
\label{Br_B2mumumunu_common_total_equation}
\textrm{Br} \left (B^-\,\to\, \ell^+ {\bar \nu}_{\ell}\, \ell^-  \ell^- \right ) \, =\, 
\textrm{Br} \left (B^-\,\to\, \ell^+ \ell^- {\bar \nu}_{\ell'}\, \ell'^- \right )\, -\,
\textrm{Br}_{interf} \left (B^-\,\to\, \ell^+ {\bar \nu}_{\ell}\, \ell^-  \ell^- \right ),
\end{eqnarray}
where 
\begin{eqnarray}
\label{Br_B2mumumunu_interf}
\textrm{Br}_{interf} \left (B^-   \right .&\to&\left . \ell^+ {\bar \nu}_{\ell}\, \ell^-  \ell^- \right ) =  \\
&=& 
\frac{\tau_{B^-}}{4 M_1}\,
\int
\sum\limits_{s_1,\, s_2,\, s_3,\, s_4}
\left (
{\cal M}_{fi}^{(1234)\,\dagger} {\cal M}_{fi}^{(1432)} + {\cal M}_{fi}^{(1432)\,\dagger} {\cal M}_{fi}^{(1234)}
\right )
d\Phi_4^{(1234)}. \nonumber
\end{eqnarray}
From (\ref{Br_B2mumumunu_interf}) it follows that in the calculation
of the interference contribution it is necessary to perform
five-dimension of numerical integration. It is necessary to use the
replacements (\ref{xij-vsakie}) in the matrix element ${\cal
  M}_{fi}^{(1432)}$.

\section{Numerical results}
\label{sec:B2lllnuNumericalResults}

To calculate the branching ratio, differential distributions, and
asymmetries, we use numerical values of the masses, lifetimes and decay
widths of the pseudoscalar and vector mesons, and matrix elements of the
CKM matrix from Ref. \cite{Tanabashi:2018oca}. The constants $f_{\rho
  (770)} = 154$ MeV and $f_{\omega (782)} = 46$ MeV were calculated in
 \cite{Melikhov:2004mk}.

Suitable parametrizations of the hadronic formfactors (\ref{hadr_matrix_ell-B2lllnu}), except the electromagnetic form factor
$V_b (q^2)$, were obtained in \cite{Melikhov:2000yu}. 
Using the generic formulae from 
\cite{Melikhov:1995xz,Melikhov:1997qk} it is possible to find the
following parametrization for the form factor $V_b (q^2)$, calculated
in the framework of the Dispersion Quark Model:

\begin{eqnarray}
\label{Vbq2}
V_b (q^2)\, =\,
\frac{1.044}{\displaystyle\left ( 1\, -\,\frac{q^2}{M_{\Upsilon}^2}\right )\,\left ( 1\, -\,0.81\,\frac{q^2}{M_{\Upsilon}^2} \right)},
\end{eqnarray}
where $M_{\Upsilon}$  mass of the $\Upsilon(1S)$ meson. The same method
allows us to obtain the values of the leptonic constants $f_{B_u} = 191$
MeV and $f_{B^*} = 183$ MeV.

We now calculate the branching ratio of the decay $B^- \to\, \mu^+ \mu^- {\bar
  \nu}_{e}\, e^-$. The natural kinematical cut of the pole by $x_{12}$  is
$x_{12\, \mathrm{min}} = \left ( 2 m_\mu / M_1 \right )^2 \approx 0.0016$.  In
this case, the numerical integration of the equation (\ref{dBr1234/dx12dx34}) by $x_{12}$ and $x_{34}$ gives:
\begin{eqnarray}
\label{Br_B2mumuenu}
\textrm{Br} \left (B^-\,\to\, \mu^+ \mu^- {\bar \nu}_e\, e^- \right )\,\approx\, 0.6\,\frac{\tau_{B^-}}{1.638\,\times\, 10^{-12}\,\, \textrm{s}}\,\,
\frac{|V_{ub}|^2}{1.55\,\times\, 10^{-5}}\,\times\, 10^{-7}.
\end{eqnarray}
The value of the branching of the $B^- \to\, \mu^+ \mu^-
{\bar \nu}_{e}\, e^-$ decay given in (\ref{Br_B2mumuenu}) is
approximately two times less than the corresponding value of $1.3
\times 10^{-7}$ from Refs.\cite{Danilina:2018uzr,Danilina:2018tsa}. This
difference is mostly due to the isotopic
coefficients $I_{\rho^0}$ and $I_\omega$ in (\ref{M_VMD}), while
decreases the contribution from the intermediate vector $\rho^0(770)$ and
$\omega(782)$ resonances to the total branching ratio by a factor 2. This
contribution is dominant, so the $\textrm{Br} \left (B^-\,\to\, \mu^+
  \mu^- {\bar \nu}_e\, e^- \right )$ increase by almost the same factor. Also the mean value of $V_{ub}$ is changed from $4.09 \times
10^{-3}$ \cite{Patrignani:2016xqp} to $3.94 \times 10^{-3}$
\cite{Tanabashi:2018oca}. A decrease of the branching by 10\%
is due to the use of the exact formula (\ref{dPhi1234}) for the
phase space.

\begin{figure}[tb]
\begin{center}
\begin{tabular}{cc}
\includegraphics[width=8.50cm]{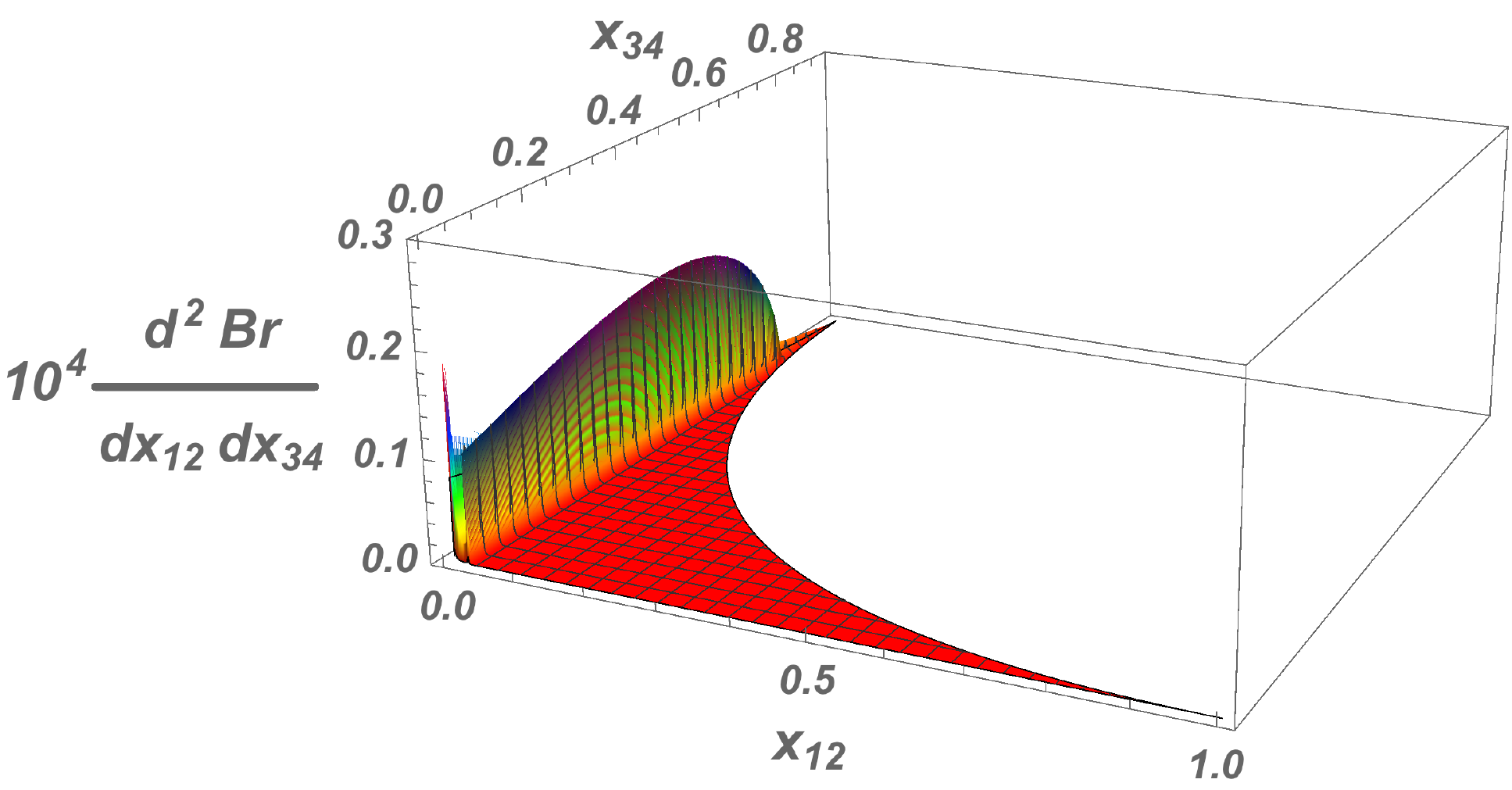} &
\includegraphics[width=8.50cm]{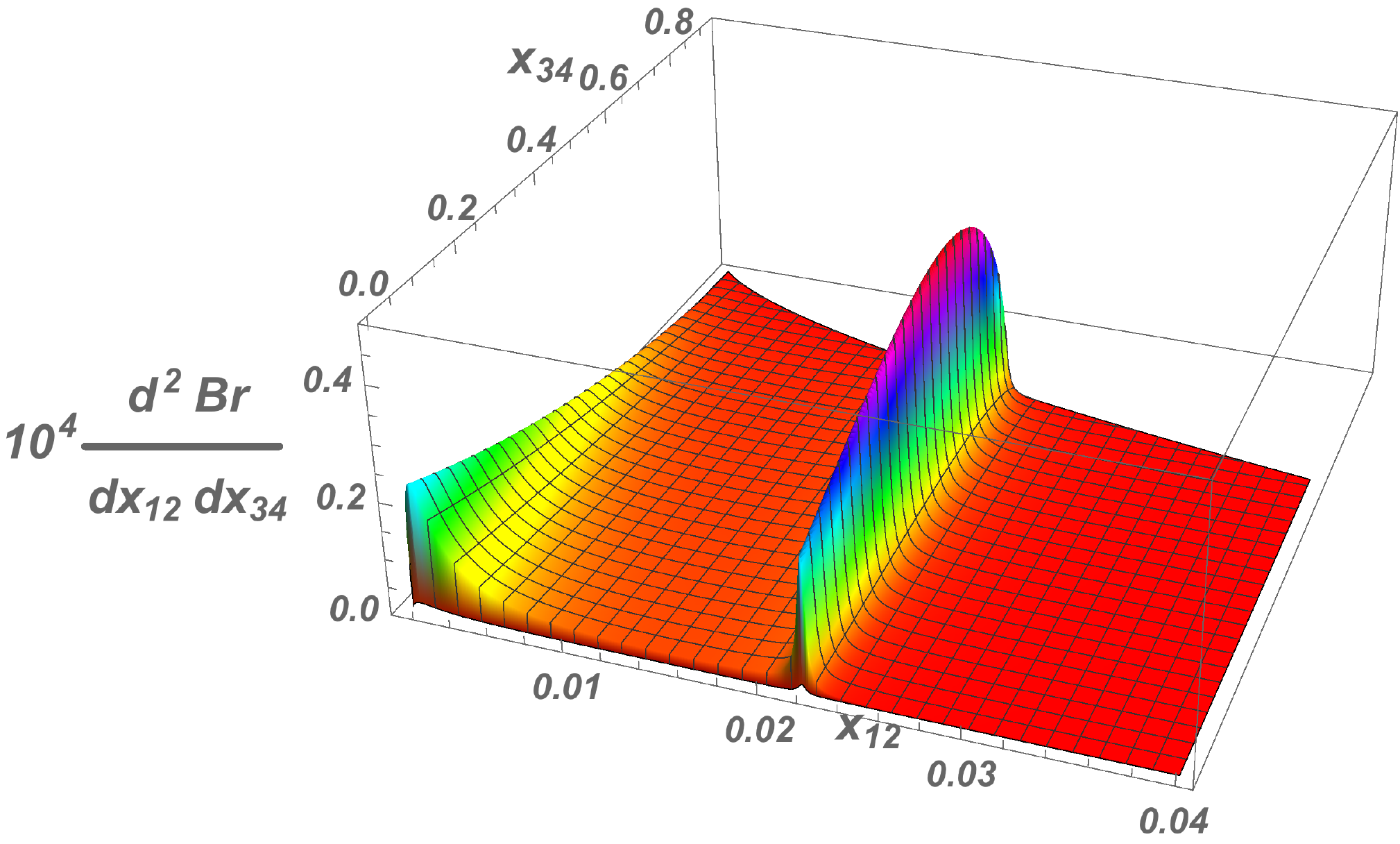} \\
\textbf{a)} & \textbf{b)} \\
\end{tabular}
\end{center}
\caption{\label{fig:dBrB2mumuenudx12dx34} Double differential
  distribution $\displaystyle 10^{4}\,\times\,\frac{d^2\textrm{Br}
    \left (B^-\,\to\, \mu^+ \mu^- {\bar \nu}_e\, e^- \right
    )}{dx_{12}\, d x_{34}}$, calculated according to formula 
  (\protect\ref{dBr1234/dx12dx34}). On Figure \textbf{b)} the range
  $x_{12} \in \left [ 0.00,\, 0.04\right ]$ is highlighted, which
  corresponds to the area of applicability of the model considered in
  the present work.}
\end{figure}

The result in equation (\ref{Br_B2mumuenu}) is compatible with the naive estimate of  (\ref{Bu2lllnu_Estimation}) up to an expected factor of two. The difference between the estimate of  (\ref{Bu2lllnu_Estimation}) and the exact
calculation (\ref{Br_B2mumuenu}) is mostly due to the fact that the
estimate does not take into account the
pole contribution when $x_{12} \to x_{12\, \mathrm{min}}$. The importance of
the pole contribution becomes obvious when analysing the double
differential distribution $d^2\,\textrm{Br} \left (B^-\,\to\,
  \mu^+ \mu^- {\bar \nu}_e\, e^- \right )/dx_{12}\, d x_{34}$, which
is presented in Fig.~(\ref{fig:dBrB2mumuenudx12dx34}). The figure
features the pole when $x_{12} \to x_{12\, \mathrm{min}} = 4 m_\mu^2/ M_1^2$
and the ridge of the narrow $\omega(782)$ resonance, the 
contribution of which defines the maximum of the matrix element. The
wide $\rho^0(770)$--meson also gives a significant contribution to the
branching ratio, but in the distribution of
Fig.~(\ref{fig:dBrB2mumuenudx12dx34}) is not a prominent as the narrow
$\omega(782)$ resonance.

The uncertainty on the numerical value of the branching ratio of the decay $B^-
\to\, \mu^+ \mu^- {\bar \nu}_{e}\, e^-$ depends on the uncertainty on
the calculation of the hadronic form factors of the transitions $B \to
\rho (770)$ and  $B \to \omega (782)$, but does not exceed 20\%
\cite{Melikhov:2000yu}. Some uncertainty is related to the use of the
Vector Meson Dominance model. This uncertainty is mostly due to
the choice of a non-perturbative phase between the summands in the
(\ref{M_VMD}).  In the VMD model this phase is equal to zero. If the
relative phase between the contributions from the $\rho^0(770)$
and $\omega(782)$ mesons into the amplitude (\ref{M_VMD}) becomes $\pi$,
then the numerical result in (\ref{Br_B2mumuenu}) may decrease to
$0.2 \times 10^{-7}$. This dependence points to the importance of a
future model-independent study of non-perturbative and non-factorized
contributions of the strong interaction to the amplitudes of the decays
$B^- \to\, \ell^+ \ell^-\, {\bar \nu}_{\ell'}\,\ell'^-$. Similar issue
of generation of additional relative phases between the contributions
of different charmonia by nonfactorizable gluons was discussed in
\cite{Lyon:2014hpa}.

In the model used for the result of  (\ref{Br_B2mumuenu}) the non-resonant
contribution, which is not related to the tails from the $\rho^0(770)$ 
and $\omega (782)$ resonances, is not taken into
account. This contribution may be estimated by using the results from Ref.
\cite{Beneke:2011nf}, in this work the branching ratio of the decay 
$B \to \gamma \ell \nu$ was predicted, omitting the contributions from
$\rho^0$ and $\omega$ resonances. An estimation of the non-resonant
contribution gives 
$$
\textrm{Br} \left (B^-\,\to\, \mu^+ \mu^- {\bar \nu}_e\, e^- \right )_{\textrm{NRC}}\, \sim \alpha_{em} \times\, \textrm{Br} \left ( B \to \gamma \ell \nu\right )_{\textrm{Beneke}}\,\sim\, 0.1\,\times\, 10^{-7},
$$
which is about 15\%  of the value of the branching ratio of  (\ref{Br_B2mumuenu}) and is comparable to the uncertainty of the
form factors calculation. Note that numerically the contributions to
(\ref{Br_B2mumuenu}) from the processes in Fig.~(\ref{fig:BstarGamma})
and Fig.~(\ref{fig:BremContrib}), which were taken into account, are
also comparable to the non-resonant contribution, which was not taken
into account.

It seems that the approximation of using only the contributions from
the lightest $\rho(770)$ and  $\omega(782)$ resonances, which is
used in this work, is not applicable if the branching ratio of the decay
$B^-\,\to\, \mu^+ \mu^- {\bar \nu}_e\, e^-$ will be measured in the
range of $\sqrt{q^2} > 1$  GeV.
In this range it is necessary to take into account the contributions
from the $\omega(1420)$, $\rho(1450)$, $\omega (1650)$, and $\rho (1700)$
resonances. These contributions should not affect the branching ratio of the
decay  $B^- \to\, \mu^+ \mu^- {\bar \nu}_{e}\, e^-$ for $\sqrt{q^2}
\le 1$ GeV but will define the behavior in the range $\sqrt{q^2} > 1$
GeV. However in the experimental procedure \cite{Aaij:2018pka}, the
variable $\sqrt{q^2}$ is chosen to be less than 980 MeV, in order to
remove a potential background from the decay $\phi \to \ell^+ \ell^-$.
So the experimental data are available only in the range of
applicability of the current work. This fact allows as to exclude from
 consideration resonances heavier than the $\rho^0 (770)$ and
$\omega (782)$.

We calculate the branching ratio of the decay $B^- \to\,e^+ e^-
{\bar \nu}_{\mu}\, \mu^-$. Formal integration in the range around the
photon pole by $x_{12}$ leads to the rough dependence of the
branching on $x_{12\,\, \mathrm{min}}$:
$$
\textrm{Br}\,\sim\,\int\,\frac{d x_{12}}{x_{12}^2}\,\sim\,\frac{1}{x_{12\,\, \mathrm{min}}}.
$$
If we choose $x_{12\,\, \mathrm{min}} =  (2 m_e/M_1)^2$, then by the order of
magnitude 
$$
\textrm{Br} \left ( B^- \to\,e^+ e^- {\bar \nu}_{\mu}\, \mu^-\right )\,\sim\,\left (\frac{m_\mu}{m_e}\right )^2\,\textrm{Br} \left (B^-\,\to\, \mu^+ \mu^- {\bar \nu}_e\, e^- \right )\,\sim\, 10^{4}\,\,\textrm{Br} \left (B^-\,\to\, \mu^+ \mu^- {\bar \nu}_e\, e^- \right ) .
$$
Because the efficiency of detection of the muonic pairs
for $\sqrt{q^2}$ below 80--100 MeV is low, this range is not
suitable for the an experimental observation. On the other hand, if we
choose $x_{12\,\, \mathrm{min}} =  (\Lambda /M_1)^2 = 0.0002$ for $\Lambda =
80$ MeV, then
\begin{eqnarray}
\label{Br_B2eemunu}
\left . \textrm{Br} \left (B^- \to\,e^+ e^- {\bar \nu}_{\mu}\, \mu^-  \right )\right |_{x_{12\,\, \mathrm{min}} =0.0002}\,\approx\, 3.0\,\frac{\tau_{B^-}}{1.638\,\times\, 10^{-12}\,\, \textrm{s}}\,\,\frac{|V_{ub}|^2}{1.55\,\times\, 10^{-5}}\,\times\, 10^{-7}.
\end{eqnarray}
The $\textrm{Br} \left (B^- \to\,e^+ e^- {\bar \nu}_{\mu}\, \mu^-
\right )$ will decrease with increasing $x_{12\,\, \mathrm{min}}$.
 
The decays $B^-\,\to\, \mu^+ \mu^- {\bar \nu}_e\, e^-$ and $B^-
\to\,e^+ e^- {\bar \nu}_{\mu}\, \mu^-$ may be suitable for tests of
the hypothesis of leptonic universality, if one measures the
branching ratio for the fixed value of $x_{12}\, >\, \left ( 2 m_\mu/ M_1
\right )^2 = 0.0016$. If leptonic universality holds, then the condition:
\begin{eqnarray}
\label{univthsalnostB2mumuenu}
\left .\frac{\textrm{Br} \left (B^-\,\to\, \mu^+ \mu^- {\bar \nu}_e\, e^- \right )}{\textrm{Br} \left ( B^- \to\,e^+ e^- {\bar \nu}_{\mu}\, \mu^-\right )}\,\right |_{x_{12}\, >\, 0.0016} =\, 1.
\end{eqnarray}
If the hints for \cite{Aaij:2014ora,Aaij:2015yra,Aaij:2017vbb,Aaij:2017uff,Aaij:2019wad} violation of the leptonic universality
 are true, then the equation (\ref{univthsalnostB2mumuenu}) may be violated.

We consider predictions for the branching ratio of the decay $B^- \to\, \mu^+
{\bar \nu}_{\mu}\, \mu^-  \mu^-$, which is the more suitable for
experimental observation \cite{Aaij:2018pka}, as the efficiency of
muon detection is higher than the efficiency of electron
detection. Numerical integration of the interference contribution (\ref{Br_B2mumumunu_interf}) for $x_{12\, \mathrm{min}}= (2 m_\mu /M_1)^2 =
0.0016$ gives 
$$
\textrm{Br}_{interf} \left ( B^- \to\, \mu^+ {\bar \nu}_{\mu}\, \mu^-  \mu^- \right )\,\approx\, -\, 0.1\,\frac{\tau_{B^-}}{1.638\,\times\, 10^{-12}\,\, \textrm{s}}\,\,\frac{|V_{ub}|^2}{1.55\,\times\, 10^{-5}}\,\times\, 10^{-7},
$$
which is comparable due to uncertainty of the strong non-perturbative
effects, the contributions from  equations (\ref{fig:BstarGamma}) and
(\ref{fig:BremContrib}) and the result with the non-resonant contribution omitted.
So we may state that in the limit of massless leptons, with a 30\%
precision from equations (\ref{Br_B2mumumunu_common_total_equation}) and
(\ref{Br_B2mumuenu}), it follows that: 
\begin{equation}
\label{Br_B2mumumunu}
\textrm{Br} \left ( B^- \to\, \mu^+ {\bar \nu}_{\mu}\, \mu^-  \mu^- \right )\,\approx\, 0.7\,\frac{\tau_{B^-}}{1.638\,\times\, 10^{-12}\,\, \textrm{s}}\,\,
\frac{|V_{ub}|^2}{1.55\,\times\, 10^{-5}}\,\times\, 10^{-7}.
\end{equation} 
This is obtained for $x_{12\, \mathrm{min}}= (2 m_\mu /M_1)^2 = 0.0016$. This
prediction is almost four times higher than the experimental upper
limit (\ref{LHCbB2mu_mu_mu_nu}), obtained in Ref. \cite{Aaij:2018pka}. What
may explain the discrepancy between the experimental result and the
theoretical prediction? First, there is quite high uncertainty of the
theoretical prediction (\ref{Br_B2mumumunu}). Second, the
value of $\textrm{Br} \left ( B^- \to\, \mu^+ {\bar \nu}_{\mu}\, \mu^-
  \mu^- \right )$ depends on the relative phase between the
contributions of the $\rho^0(770)$ and  $\omega(782)$ resonances. In
the framework of VMD it is zero, however various non-perturbative
contributions may lead to non-zero value. All the other contribution,
which were omitted in the current work could not significantly
influence the numerical result of  equation (\ref{Br_B2mumumunu}). It seems
unlikely that the discrepancy between the prediction and
measured result may be attributed to Beyond the Standard Model physics.

The decays $B^- \to\, e^+ {\bar \nu}_{e}\, e^-  e^-$ and $B^- \to\,
\mu^+ {\bar \nu}_{\mu}\, \mu^-  \mu^-$ allow us to introduce yet another
test for lepton universality:
\begin{eqnarray}
\label{univthsalnostB2mumumunu}
\left .\frac{\textrm{Br} \left (B^- \to\, \mu^+ {\bar \nu}_{\mu}\, \mu^- \mu^-  \right )}{\textrm{Br} \left ( B^- \to\,e^+ {\bar \nu}_{e}\, e^- e^-\right )}\,\right |_{x_{12} > 0.0016,\,\,\, x_{14} > 0.0016 } =\, 1.
\end{eqnarray}

\begin{figure}[tb]
\begin{center}
\begin{tabular}{cc}
\includegraphics[width=8.75cm]{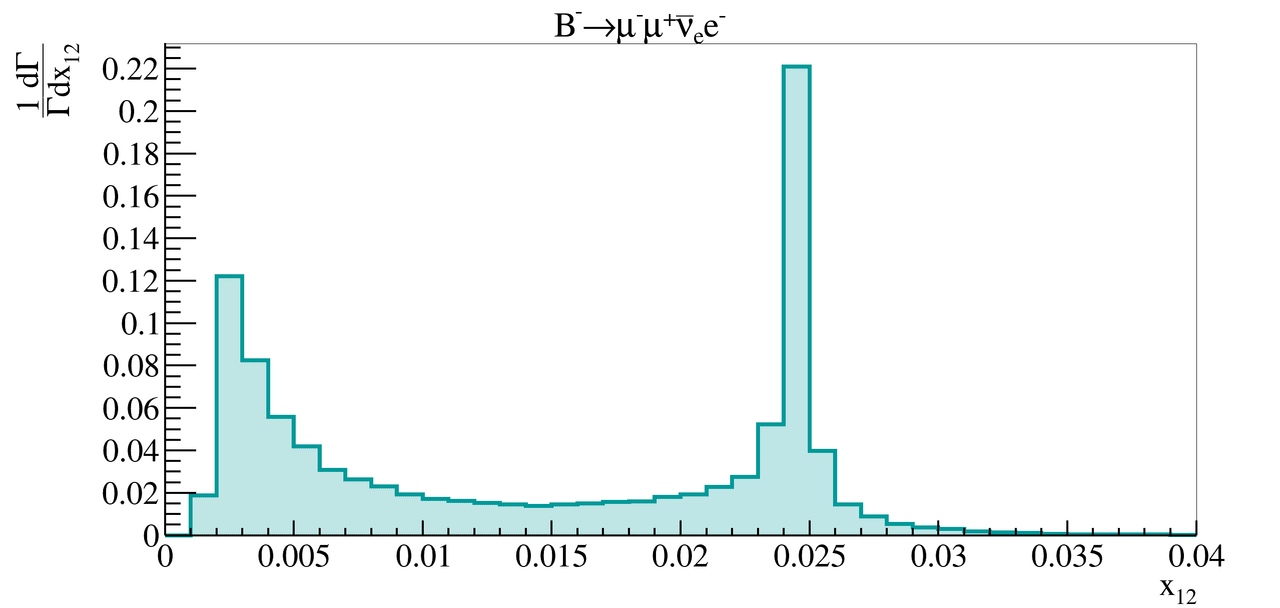} &
\includegraphics[width=8.75cm]{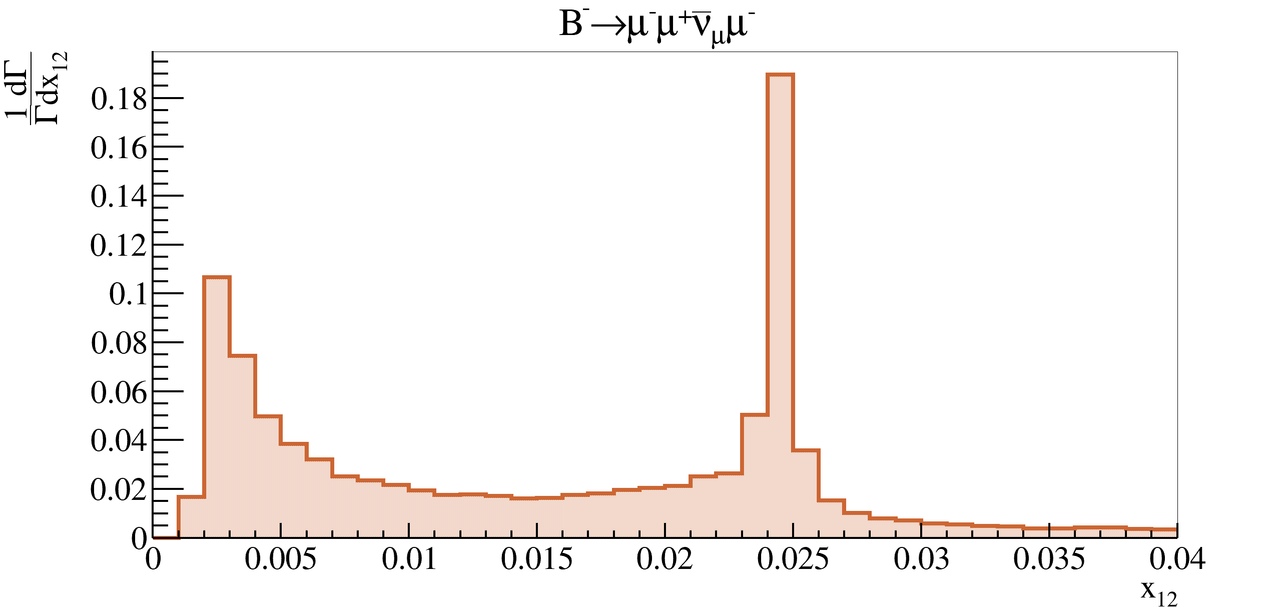} \\
\textbf{a)} & \textbf{b)} \\
\end{tabular}
\end{center}
\caption{\label{fig:dBrB2mumuenudx12} Normalized differential
  distributions $\displaystyle
  \frac{1}{\Gamma}\,\frac{d\Gamma}{dx_{12}}$ for the decays
  \textbf{a)} $B^-\,\to\, \mu^+ \mu^- {\bar \nu}_e\, e^- $ and
  \textbf{b)} $B^- \to\, \mu^+ {\bar \nu}_{\mu}\, \mu^-  \mu^-$,
  obtained by integration by $d x_{34}\, d y_{12}\, d y_{34}\, d
  \varphi$ of  equations (\protect\ref{dBr1234common})  and
  (\protect\ref{dBr_common}) respectively.}
\end{figure}

\begin{figure}[tb]
\begin{center}
\begin{tabular}{cc}
\includegraphics[width=8.75cm]{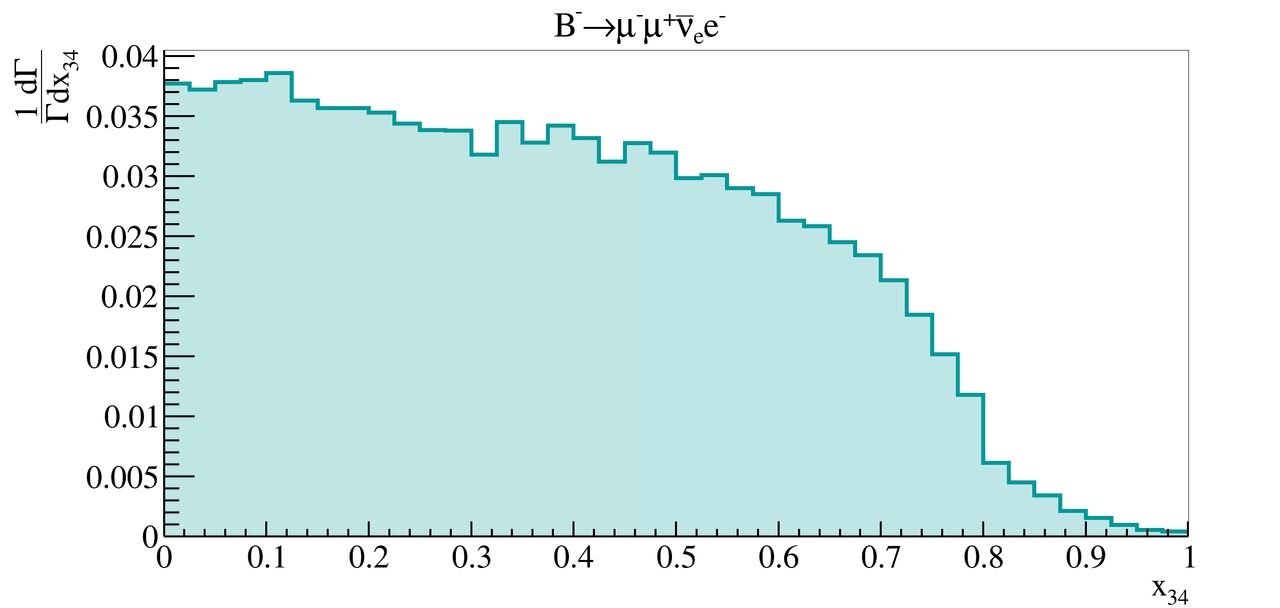} &
\includegraphics[width=8.75cm]{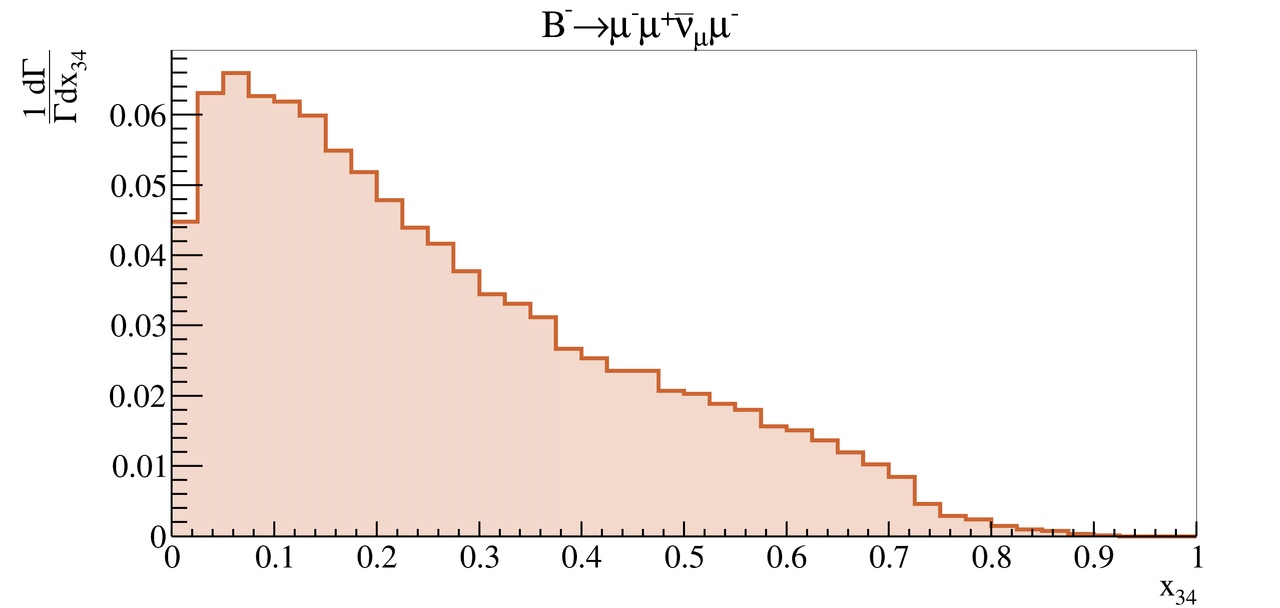} \\
\textbf{a)} & \textbf{b)} \\
\end{tabular}
\end{center}
\caption{\label{fig:dBrB2mumuenudx34} Normalized differential
  distributions $\displaystyle
  \frac{1}{\Gamma}\,\frac{d\Gamma}{dx_{34}}$ for the decays
  \textbf{a)} $B^-\,\to\, \mu^+ \mu^- {\bar \nu}_e\, e^- $  and
  \textbf{b)} $B^- \to\, \mu^+ {\bar \nu}_{\mu}\, \mu^-  \mu^-$,
  obtained by integration by $d x_{12}\, d y_{12}\, d y_{34}\, d
  \varphi$ of  formulae (\protect\ref{dBr1234common}) and
  (\protect\ref{dBr_common}) respectively.}
\end{figure}

\begin{figure}[tb]
\begin{center}
\begin{tabular}{cc}
\includegraphics[width=8.75cm]{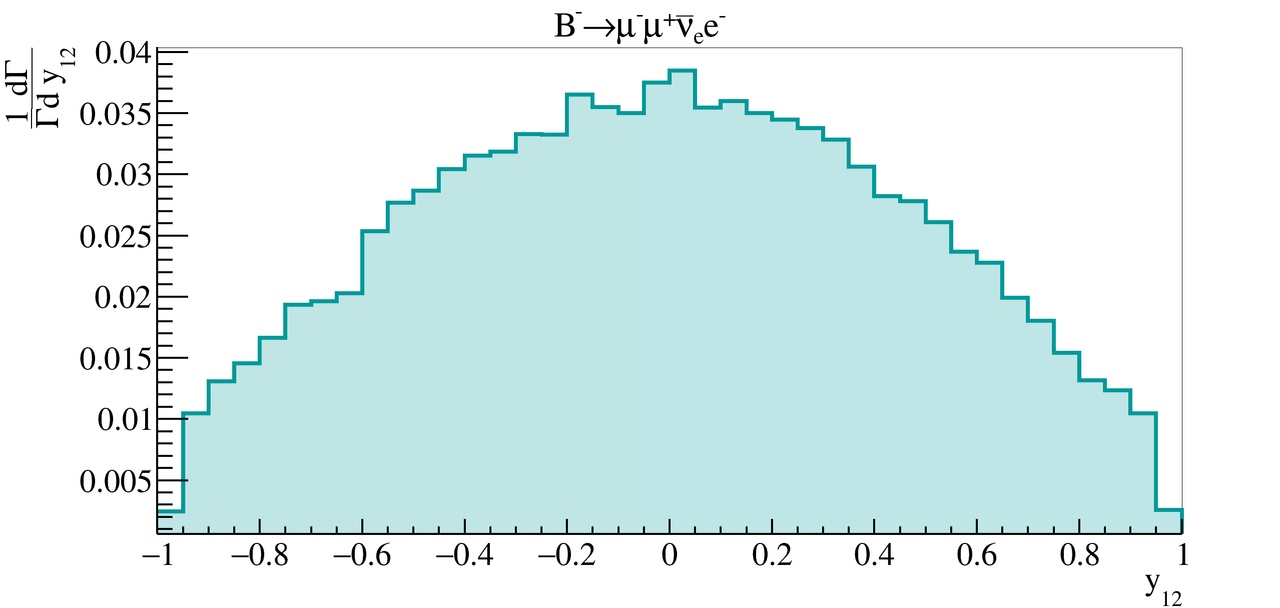} &
\includegraphics[width=8.75cm]{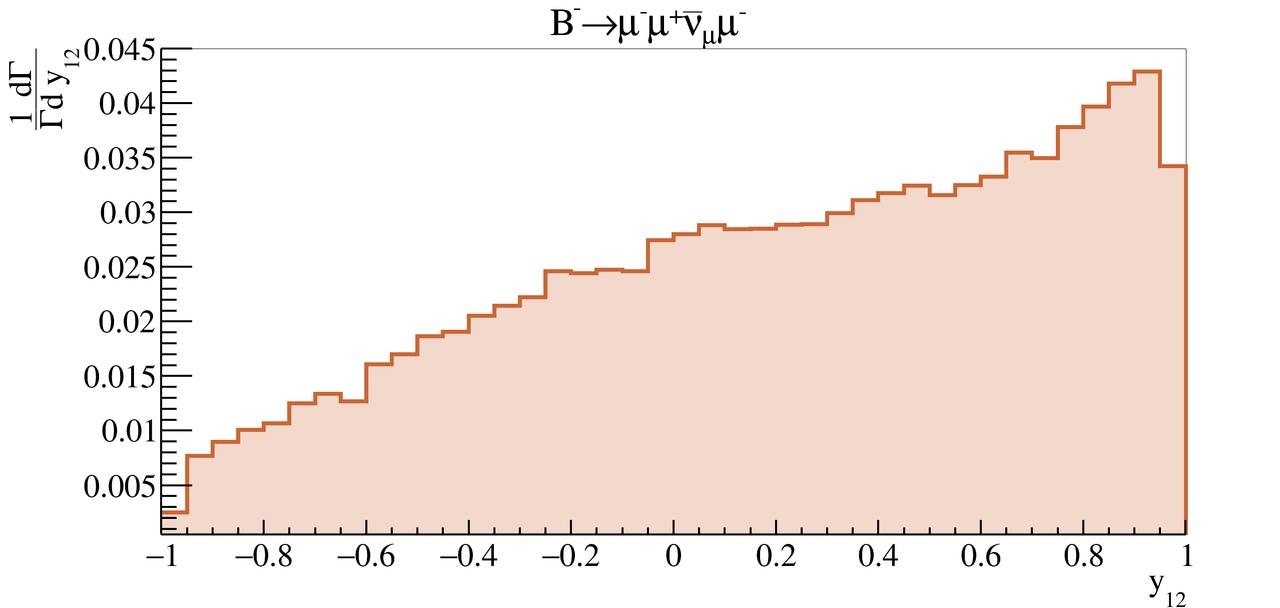} \\
\textbf{a)} & \textbf{b)} \\
\end{tabular}
\end{center}
\caption{\label{fig:dBrB2mumuenudy12} Normalized differential
  distributions $\displaystyle
  \frac{1}{\Gamma}\,\frac{d\Gamma}{dy_{12}}$ for the decays
  \textbf{a)} $B^-\,\to\, \mu^+ \mu^- {\bar \nu}_e\, e^- $  and
  \textbf{b)} $B^- \to\, \mu^+ {\bar \nu}_{\mu}\, \mu^-  \mu^-$,
  obtained by integrating by $d x_{12}\, d x_{34}\, d y_{34}\, d
  \varphi$ of formulae (\protect\ref{dBr1234common}) and
  (\protect\ref{dBr_common}) accordingly.}
\end{figure}

We consider single differential distributions for the decays $B^-
\to\, \mu^+ \mu^- {\bar \nu}_{e}\, e^-$ and $B^- \to\, \mu^+ {\bar
  \nu}_{\mu}\, \mu^-  \mu^-$.
One-dimensional projections of the double differential distribution
$\displaystyle\frac{d^2\textrm{Br}}{dx_{12}\, d x_{34}}$ by $x_{12}$
and $x_{34}$ are given in
Fig.~(\ref{fig:dBrB2mumuenudx12}) and~(\ref{fig:dBrB2mumuenudx34})
respectively. The distributions by $x_{12}$ are given in the range
$\left [0,\, 0.04 \right ]$, which corresponds to the area of
applicability of the model. Fig.~(\ref{fig:dBrB2mumuenudx12})
features a photon pole for $x_{12} \to x_{12\, \mathrm{min}}= (2 m_\mu /M_1)^2 =
0.0016$ and a peak from the $\omega(782)$ resonance for $x_{12} \to
(M_\omega  /M_1)^2 \approx 0.023$. Due to the fact that the $\rho^0
(770)$ meson has a width of about 150 MeV, the contribution from this
meson in Fig.~(\ref{fig:dBrB2mumuenudx12}) appears as a wide background
to the narrow peak of the $\omega(782)$ resonance. The distributions
by $x_{34}$ in Fig.~(\ref{fig:dBrB2mumuenudx34}) does not have poles, in agreement with the analysis from
Section~\ref{sec:strukturaB2lllnu}, and 
demonstrates the importance of taking into account the Fermi
antisymmetry in the decay $B^- \to\, \mu^+ {\bar \nu}_{\mu}\, \mu^-
\mu^-$, because due to the additional contribution from Fermi
antisymmetry the shapes of the distributions by $x_{34}$
in the decays $B^- \to\, \mu^+ \mu^- {\bar \nu}_{e}\, e^-$ and $B^- \to\,
\mu^+ {\bar \nu}_{\mu}\, \mu^-  \mu^-$ are significantly different.
An analogous difference may be seen in the distributions by $y_{12} =
\cos\theta_{12}$ and $y_{34} = \cos\theta_{34}$,
which are presented in Fig.~(\ref{fig:dBrB2mumuenudy12})
and (\ref{fig:dBrB2mumuenudy34}) respectively. The definition of
angular variables $y_{12}$ and $y_{34}$ is given in
Appendix~\ref{sec;kinemat4}.

\begin{figure}[tb]
\begin{center}
\begin{tabular}{cc}
\includegraphics[width=8.75cm]{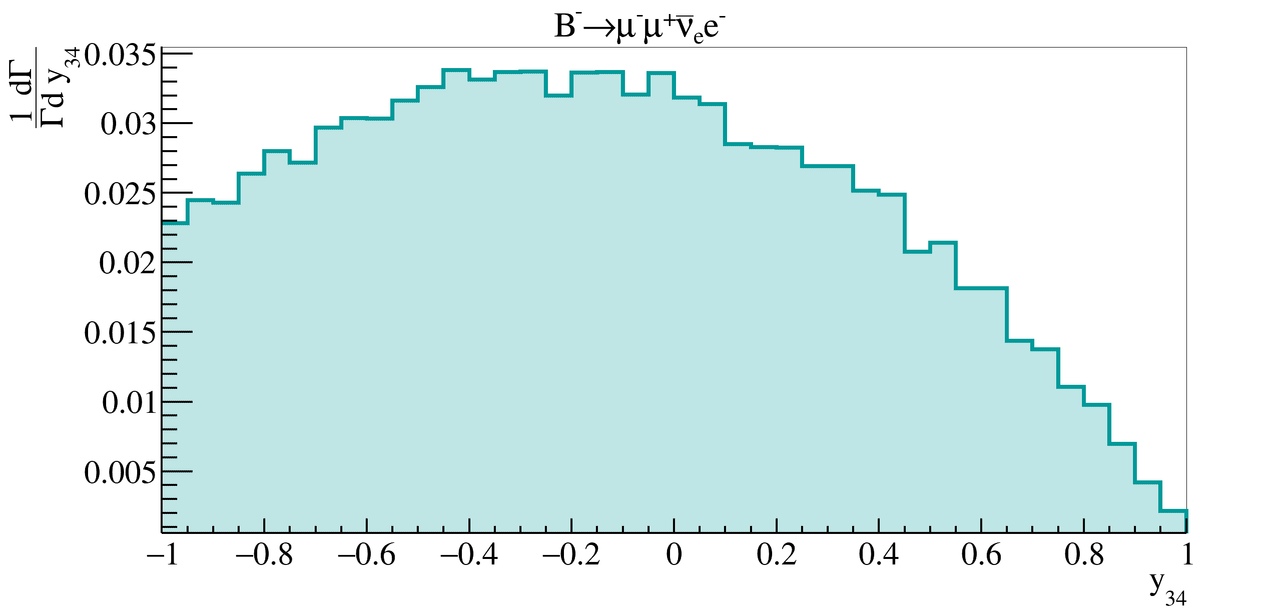} &
\includegraphics[width=8.75cm]{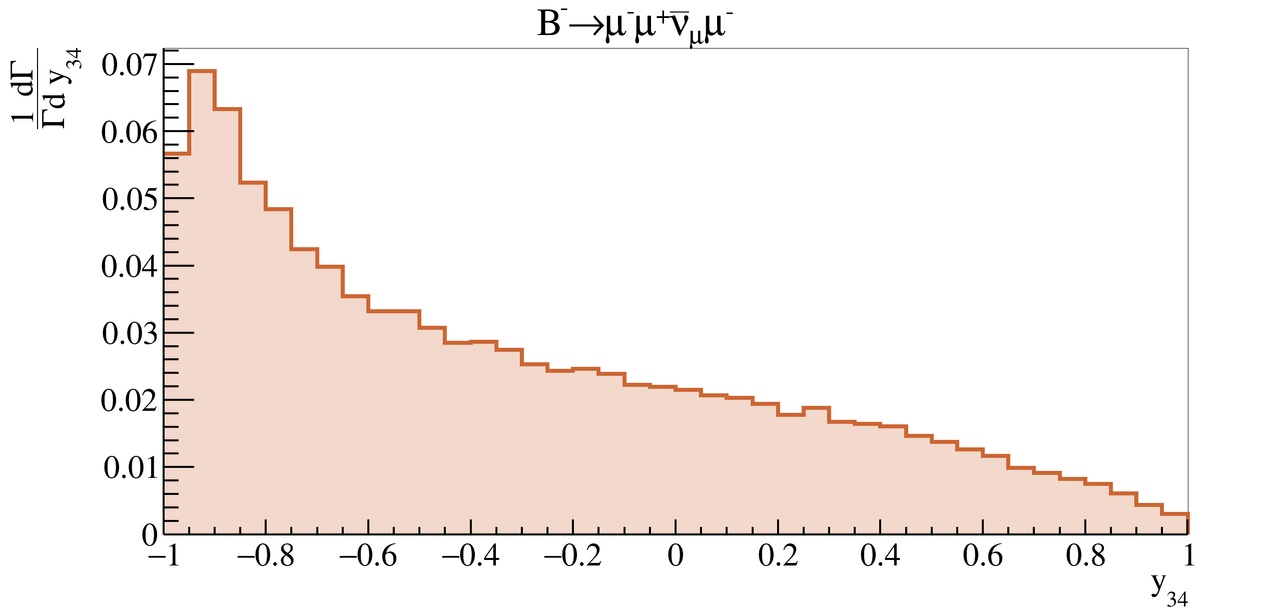} \\
\textbf{a)} & \textbf{b)} \\
\end{tabular}
\end{center}
\caption{\label{fig:dBrB2mumuenudy34} Normalized differential
  distributions $\displaystyle
  \frac{1}{\Gamma}\,\frac{d\Gamma}{dy_{34}}$ for the decays
  \textbf{a)} $B^-\,\to\, \mu^+ \mu^- {\bar \nu}_e\, e^- $  and
  \textbf{b)} $B^- \to\, \mu^+ {\bar \nu}_{\mu}\, \mu^-  \mu^-$,
  obtained by integration by $d x_{12}\, d x_{34}\, d y_{12}\, d
  \varphi$ of  equations  (\protect\ref{dBr1234common}) and
  (\protect\ref{dBr_common})  respectively.}
\end{figure}

\begin{figure}[tb]
\begin{center}
\begin{tabular}{cc}
\includegraphics[width=8.75cm]{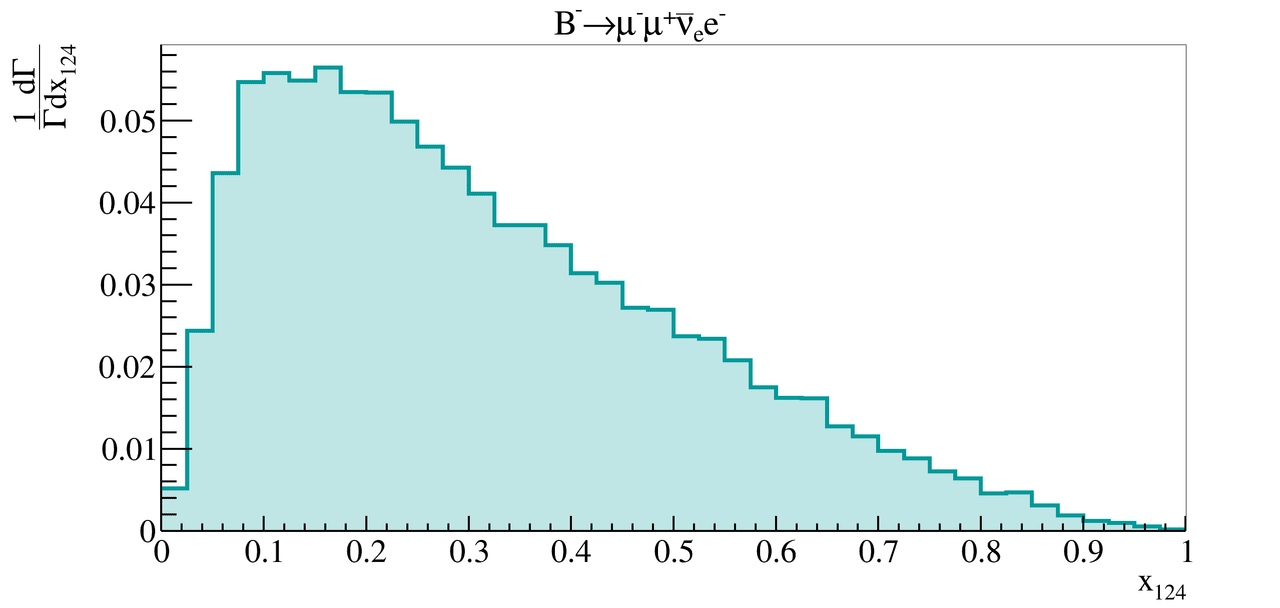} &
\includegraphics[width=8.75cm]{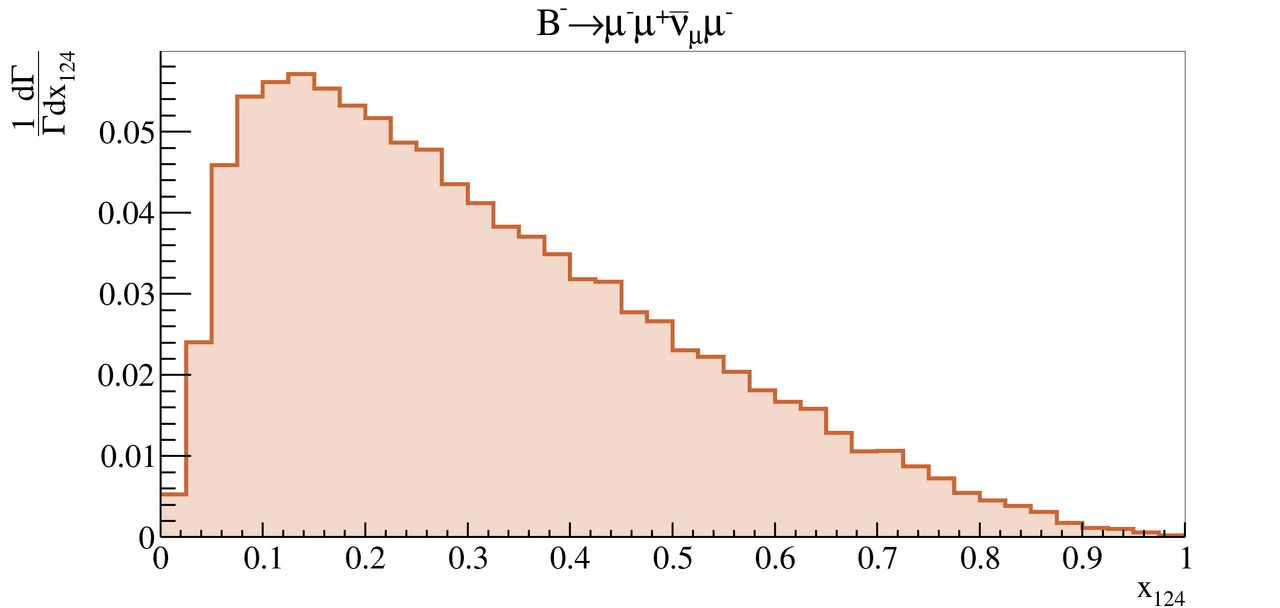} \\
\textbf{a)} & \textbf{b)} \\
\end{tabular}
\end{center}
\caption{\label{fig:dBrB2mumuenudx124} Normalized differential
  distributions $\displaystyle
  \frac{1}{\Gamma}\,\frac{d\Gamma}{dx_{124}}$ by invariant mass of all
  of the charged leptons in the final state for the decays \textbf{a)}
  $B^-\,\to\, \mu^+ \mu^- {\bar \nu}_e\, e^- $  and \textbf{b)} $B^-
  \to\, \mu^+ {\bar \nu}_{\mu}\, \mu^-  \mu^-$.}
\end{figure}

Detectability of the multi-lepton decays of the $B$--mesons with a neutrino in the final
state may be linked to the distributions by normalized invariant mass of the charged
leptons. The square of the corresponding mass is defined as:
\begin{eqnarray}
\label{x124-def}
x_{124}\, =\,\frac{(k_1 + k_2 + k_4)^2}{M_1^2},
\end{eqnarray}
where the $k_i$ are four-momenta of charged leptons in the final
state. The distributions by $x_{124}$ are presented in
Fig.~(\ref{fig:dBrB2mumuenudx124}). One can see from the figure that the
shape of the distribution by $x_{12}$ is not very sensitive to the
procedure of Fermi antisymmetrization.

\begin{figure}[tb]
\begin{center}
\begin{tabular}{cc}
\includegraphics[width=8.50cm]{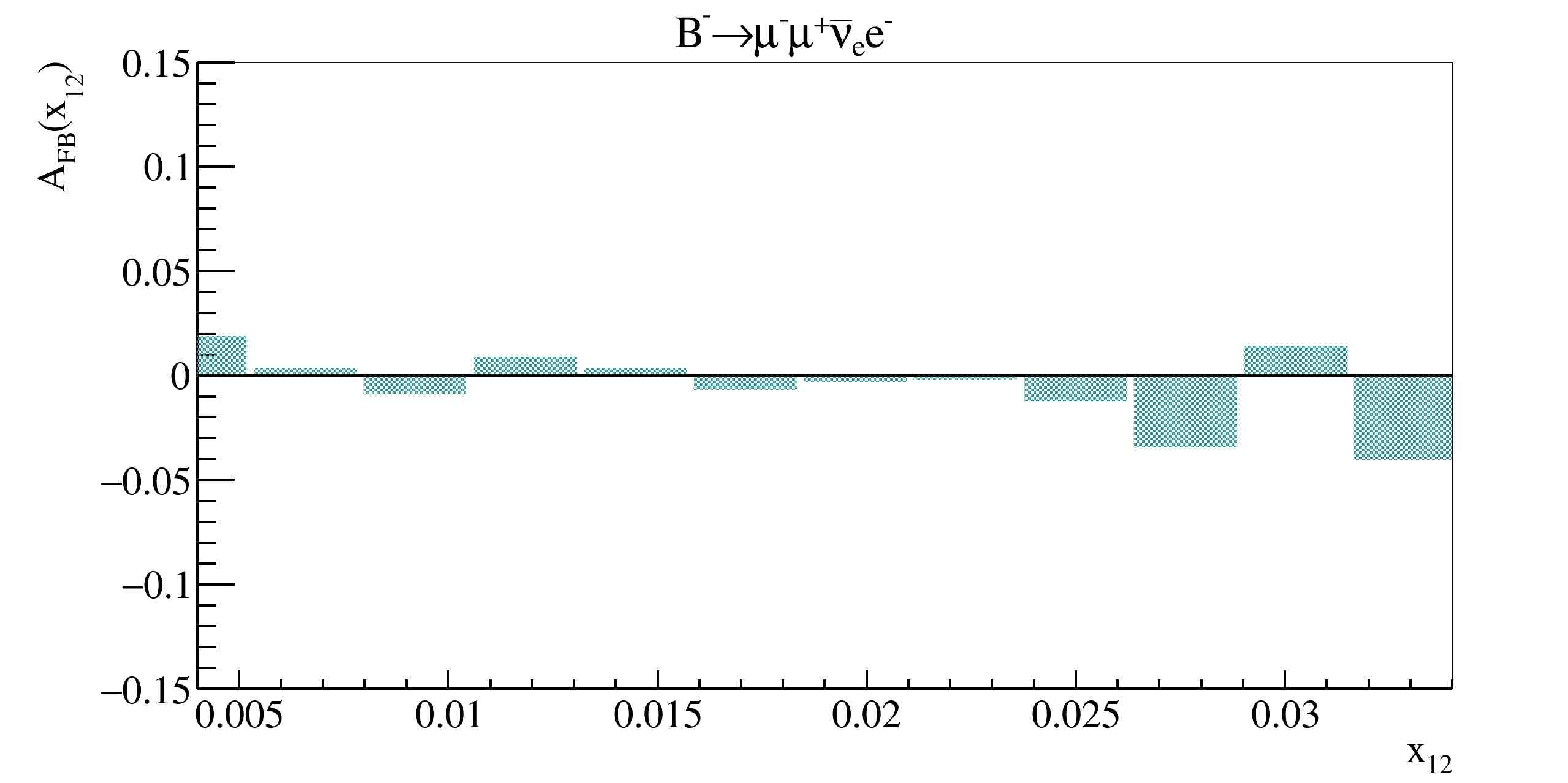} &
\includegraphics[width=8.80cm]{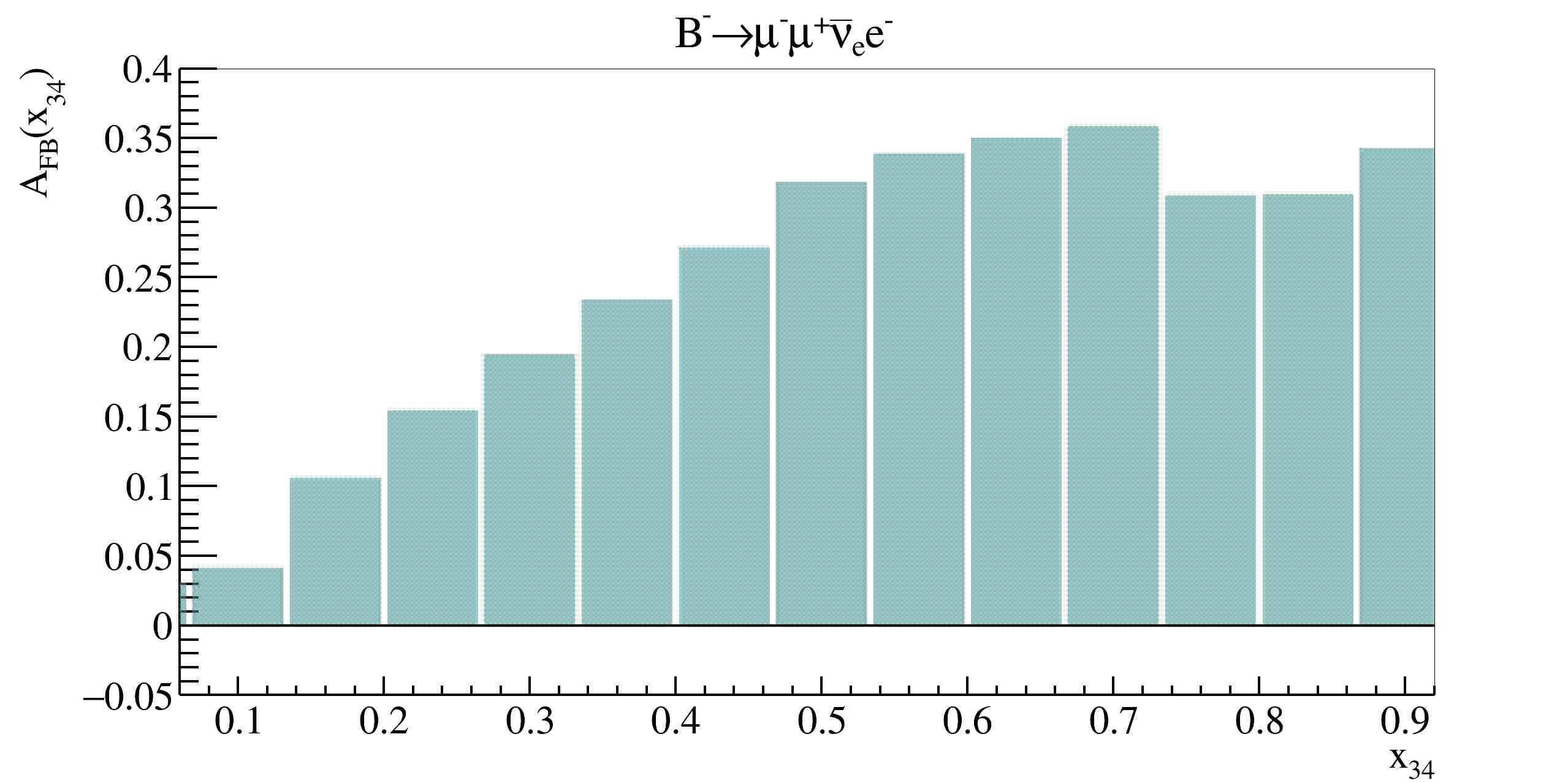} \\
\textbf{a)} & \textbf{b)} \\
\end{tabular}
\end{center}
\caption{\label{fig:dBrB2mumuenudxA_FB} Forward--backward lepton asymmetries
  \textbf{a)} $A^{(B^-)}_{FB}(x_{12})$ and \textbf{b)}
  $A^{(B^-)}_{FB}(x_{34})$ for the decay $B^-\,\to\, \mu^+ \mu^- {\bar
    \nu}_e\, e^- $, calculated using equations (\protect\ref{Afbq2-def}) and (\protect\ref{Afbk2-def})
  respectively.}
\end{figure}

It is well known that forward--backward lepton asymmetries are very
sensitive to BSM physics. For the decay $B^-\,\to\, \mu^+ \mu^- {\bar
  \nu}_e\, e^- $ it is possible to define forward--backward lepton
asymmetries $A^{(B^-)}_{FB}(x_{12})$ and $A^{(B^-)}_{FB}(x_{34})$
according to equations (\ref{Afbq2-def}) and (\ref{Afbk2-def}). These
asymmetries are shown in Fig.~(\ref{fig:dBrB2mumuenudxA_FB}). The
asymmetry $A^{(B^-)}_{FB}(x_{12})$ is shown only for the interval
$x_{12} \in \left [0,\, 0.04 \right ]$, which corresponds to the area
of applicability of the current model. In this interval,
excluding the area of the $\omega(782)$ resonance,
the contributions to $A^{(B^-)}_{FB}(x_{12})$  come from
electromagnetic and strong processes;thus this asymmetry is close to zero
in almost all of the considered range. The shape of the asymmetry
$A^{(B^-)}_{FB}(x_{34})$ is very similar to the shape of the
asymmetries in three-body semileptonic decays of $B$--mesons.

One cannot to study forward--backward lepton asymmetries in the decay
$B^- \to\, \mu^+ {\bar \nu}_{\mu}\, \mu^-  \mu^-$, as in this case
there are two identical negative muons in the final state. Experimentally it
is not possible to distinguish which of the negatively charged muons
should be attributed to the $\mu^+ \mu^-$--pair, and which to the
${\bar \nu}_{\mu}\, \mu^-$--pair. 

All the above  that is related to the differential distributions for the decays
$B^-\,\to\, \mu^+ \mu^- {\bar \nu}_e\, e^- $ and $B^- \to\, \mu^+
{\bar \nu}_{\mu}\, \mu^-  \mu^-$ is also related to the differential
distributions for the decays $B^-\,\to\, e^+ e^- {\bar \nu}_\mu\,
\mu^- $ and $B^- \to\, e^+ {\bar \nu}_{e}\, e^-  e^-$. In this model the lepton universality holds, so the differential
distributions of the two latter decays are not needed.

\section*{Conclusion}

In the present work,
\begin{itemize}
\item theoretical predictions for the branching ratios of the decays
  $B^-\,\to\, \mu^+ \mu^- {\bar \nu}_e\, e^- $ and $B^- \to\, \mu^+
  {\bar \nu}_{\mu}\, \mu^-  \mu^-$ are obtained in the framework of Standard Model:
 \begin{eqnarray}
\textrm{Br} \left (B^-\,\to\, \mu^+ \mu^- {\bar \nu}_e\, e^- \right )\,\approx\, 0.6\,\times\, 10^{-7} \nonumber
\end{eqnarray}
and
\begin{equation}
\textrm{Br} \left ( B^- \to\, \mu^+ {\bar \nu}_{\mu}\, \mu^-  \mu^- \right )\,\approx\, 0.7\,\times\, 10^{-7}, \nonumber
\end{equation} 
and uncertainties for every prediction are discussed; 
\item the difference between the obtained predictions and the
  predictions from Ref.\cite{Danilina:2018uzr} is discussed, as well as
  the compatibility with the recent experimental result \cite{Aaij:2018pka} by the LHCb
  collaboration;
\item the possibility to test the hypothesis of lepton universality in 
  rare four-leptonic decays of $B$--mesons with three charged leptons
  in the final state is analysed;
\item double and single differential distributions for the decays $B^-\,\to\, \mu^+ \mu^- {\bar
    \nu}_e\, e^- $ and $B^- \to\, \mu^+ {\bar \nu}_{\mu}\, \mu^-
  \mu^-$ are considered, and some recommendations for searches for Beyond the Standard Model 
  physics in these decays are given.
\end{itemize}


\section*{Acknowledgements}
The authors would like to thank I.~M.~Belyaev (ITEP), E.~E.~Boos (SINP
MSU), L.~V.~Dudko (SINP MSU), V.~Yu.~Yegorychev (ITEP),
A.~D.~Kozachuk (SINP MSU), and  D.~V.~Savrina (ITEP, SINP MSU) for
fruitful discussions which improved the current work significantly.

The authors would like to especially thank D.~I.~Melikhov (SINP MSU)
for help with calculation of form factor $V_b (q^2)$  and numerous
fruitful discussions.

The authors would like to express their deep gratitude to Prof.~Sally
Seidel (UNM, USA) for help with preparation of the paper.

The work was supported by grant 16-12-10280 of the Russian Science
Foundation. The authors (A.~Danilina and N.~Nikitin)
express their gratitude for this support.

A.~Danilina is grateful to the ``Basis'' Foundation for her stipend for
Ph.D. students.

\appendix

\section{Kinematics of four-lepton decays}
\label{sec;kinemat4}

Denote the four-momenta of  the final leptons in four-leptonic decays
of $B$--mesons as $k_i$, $i =\{1,\, 2,\, 3,\, 4\}$.  Let 
$$
q = k_1 + k_2; \quad
k = k_3 + k_4; \quad
\tilde q = k_1 + k_4; \quad
\tilde k = k_2 + k_3; \quad
p = k_1 + k_2 + k_3 + k_4,
$$ 
where $p$ is the four-momentum of the $B$--meson and $p^2 = M^2_1$. For the
calculations below it is suitable to use the dimensionless
variables: 
$$
x_{12} = \frac{q^2}{M_1^2}, \quad
x_{34} = \frac{k^2}{M_1^2}, \quad
x_{14} = \frac{{\tilde q}^2}{M_1^2}, \quad
x_{23} = \frac{{\tilde k}^2}{M_1^2}.
$$
By common notation, $x_{ij} = (k_i + k_j)^2/M^2_1$. Hence $x_{ij} =
x_{ji}$. The leptons may be considered as massless in almost all of
the calculations of the present work, i.e., $k_i^2=0$. However during
the calculation of the bremsstrahlung contribution in the area $q^2
\sim {4m_\ell}^2$, where $m_\ell$ is the mass of any of the charged
leptons of the $\ell^+\ell^-$ pair, it is necessary to take into account
the dependence of the bremsstrahlung matrix element and phase space on
the value of $m_\ell$.

From the conservation law of four-momentum that in the
zero-mass limit the variables $x_{ij}$ are linked by 
\begin{eqnarray}
\label{monentum-equation}
x_{12} + x_{13} + x_{14} + x_{23} + x_{24} + x_{34} = 1.
\end{eqnarray}
Let us find the intervals for $x_{ij}$ using the inequality
$(p_1 p_2) \ge \sqrt{p_1^2\, p_2^2}$; then any $x_{ij} \ge 0$. On the
other hand,
\begin{eqnarray}
1 & =&\frac{p^2}{M_1^2}\, =\,\frac{(q+k)^2}{M^2_1}\,\ge \,\frac{(\sqrt{q^2} + \sqrt{k^2})^2}{M_1^2}\, =\,\Big (\sqrt{x_{12}} + \sqrt{x_{34}} \Big)^2.
\nonumber
\end{eqnarray}
As $0 \le x_{34}$, then  $x_{12} \le 1$, so $x_{12} \in [0,\, 1]$. The
upper limit of the variable $x_{34}$ depends on the value of $x_{12}$:
$$
x_{34}\, =\,\frac{(p - q)^2}{M_1^2}\,\le\,\frac{(M_1 - \sqrt{q^2})^2}{M_1^2}\, =\,\left ( 1 - \sqrt{x_{12}}\right )^2.
$$
Thus for a fixed value of $x_{12}$ the variable $x_{34} \in \left [ 0,\,\left ( 1 - \sqrt{x_{12}}\right )^2 \right ]$.

For the pair $x_{14}$ and $x_{23}$, the analogous condition holds: $x_{14} \in
[0,\, 1]$ and for a fixed $x_{14}$, $x_{23} \in \left [ 0,\,\left ( 1 - \sqrt{x_{14}}\right )^2 \right ]$.

\begin{figure}[bt]
\begin{center}
\includegraphics[width=11.0cm]{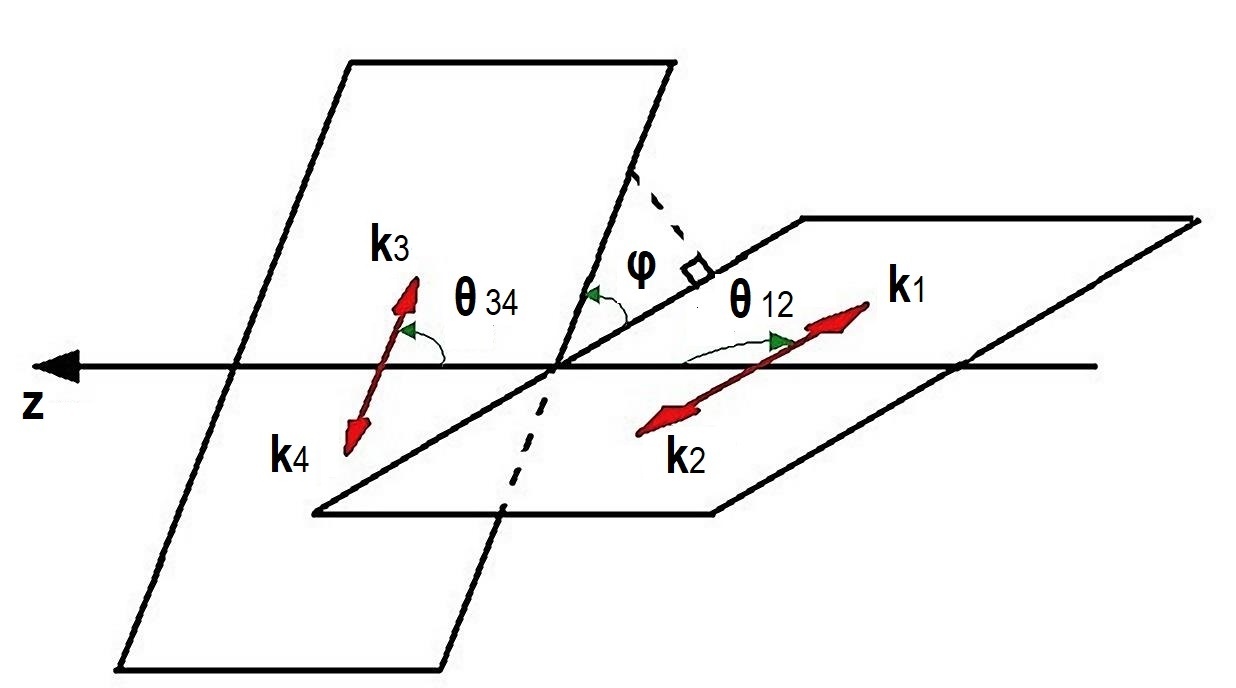} 
\end{center}
\caption{\protect\label{fig:kinematika} 
Kinematics of the decay $B^-(p) \,\to\, \ell^+(k_1)\,\ell^-(k_2)\,
{\bar \nu}_{\ell'}(k_3)\,\ell'^-(k_4)$.  Angle $\theta_{12}$ is
defined in the rest frame of $\ell^+(k_1)\,\ell^-(k_2)$--pair;  angle
$\theta_{34}$ is defined in the rest frame of ${\bar
  \nu}_{\ell'}(k_3)\,\ell'^-(k_4)$--pair; angle $\varphi$ is defined
in the rest frame of $B^-$--meson.
}
\end{figure}

Consider the kinematics of the decay
$
B^-(p) \,\to\, \ell^+(k_1)\,\ell^-(k_2)\, {\bar \nu}_{\ell'}(k_3)\,\ell'^-(k_4) 
$,
when the flavor of negatively charged lepton $\ell^-(k_2)$ is
different from the flavor of the negatively charged lepton
$\ell'^-(k_4)$. 
Let the positively charged lepton have the momentum ${\bf k}_1$, and let the 
antineutrino have the momentum ${\bf k}_3$. We define an angle
$\theta_{12}$ between the momentum of the positively charged lepton and
the direction of the $B$--meson ($z$--axis) in the rest frame of the
$\ell^+\ell^-$ pair, and another angle $\theta_{34}$ between the
direction of the antineutrino and the direction of the $B$--meson
($z$--axis) in the rest frame of $\ell'^- \bar\nu_{\ell'}$--pair. Then
\begin{eqnarray}
\label{costheta-equation}
y_{12}\,\equiv\,\cos\theta_{12} &=& \frac{1}{\lambda^{1/2} (1,\, x_{12},\, x_{34})}\,\left ( x_{23} + x_{24} - x_{13} - x_{14}\right ),
\\
y_{34}\,\equiv\,\cos\theta_{34} &=& \frac{1}{\lambda^{1/2} (1,\, x_{12},\, x_{34})}\,\left ( x_{14} + x_{24} - x_{13} - x_{23}\right ),
\nonumber
\end{eqnarray}
where $\lambda (a, \, b,\, c) = a^2 + b^2 + c^2 - 2ab - 2ac - 2bc$, the 
triangle function. Angles $\theta_{12} \in \left [0,\,\pi \right ]$
and $\theta_{34} \in \left [0,\,\pi \right ]$. Hence $y_{12} \in
[-1,\, 1]$ and $y_{34} \in [-1,\, 1]$. Angles are measured relative to
$z$--axis. Also let us define an angle $\varphi \in [0,\, 2 \pi)$ in
the rest frame of the $B$--meson between the planes which are set by the pairs
of vectors $({\bf k}_1,\, {\bf k}_2)$ and $({\bf k}_3,\, {\bf
  k}_4)$.  Introduce a vector ${\bf  a}_1 = {\bf k}_1\,\times\,
{\bf k}_2$, perpendicular to the plane $({\bf k}_1,\, {\bf k}_2)$, and 
vector ${\bf  a}_3 = {\bf k}_{\, 4}\,\times\, {\bf k}_{\, 3}$,
perpendicular to the plane $({\bf k}_3,\, {\bf k}_4)$. Then
$$
\cos\varphi\, =\,\frac{\Big ( {\bf  a}_1,\, {\bf  a}_3\Big )}{|{\bf  a}_1|\, |{\bf  a}_3|}.
$$
Using the technique from Ref. \cite{BuKa}, for $\cos\varphi$ we can write:
\begin{eqnarray}
\label{cosvarphi-3lnu}
\cos\varphi\, =\,\frac{\textrm{det}\,
\left (
\begin{matrix}
M_1^2   & (p\, k_1) & (p\, k_2) \\
(p\, k_4) & (k_1\, k_4) & (k_2\, k_4) \\
(p\, k_3) & (k_1\, k_3) & (k_2\, k_3)
\end{matrix}
\right )}{\sqrt{
\textrm{det}\,
\left (
\begin{matrix}
M_1^2   & (p\, k_1) & (p\, k_2) \\
(p\, k_1) & 0 & (k_1\, k_2) \\
(p\, k_2) & (k_1\, k_2) & 0
\end{matrix}
\right )\,\,
\textrm{det}\,
\left (
\begin{matrix}
M_1^2   & (p\, k_3) & (p\, k_4) \\
(p\, k_3) & 0 & (k_3\, k_4) \\
(p\, k_4) & (k_3\, k_4) & 0
\end{matrix}
\right )
}}.
\end{eqnarray}
Simplifying the (\ref{cosvarphi-3lnu}) gives
\begin{eqnarray}
\label{cosvarphi-equation}
-\,2\,\sqrt{x_{12}\, x_{34}\, (1 - y^2_{12})\, (1 - y^2_{34})}\,\cos\varphi\, +\, (1 - x_{12} - x_{34})\, y_{12}\, =\, x_{13} - x_{14} - x_{23} + x_{24}.
\end{eqnarray}
Four-particle phase space has the form:
$$
d\Phi_4^{(1234)}\, =\, M_1^4\,
\frac{d x_{12}}{2 \pi}\,\frac{d x_{34}}{2 \pi}\, d\Phi_2^{(qk)}\, d\Phi_2^{(12)} \, d\Phi_2^{(34)}, 
$$
where (assuming zero masses for leptons  $\ell^{\pm}$ and $\ell'^{\, -}$ masses)
we can write
\begin{eqnarray}
d\Phi_2^{(qk)} &=& 2 \pi \delta \left ( q^2 - x_{12} M_1^2 \right )\, \frac{d^4 q}{(2 \pi)^4}\,
2 \pi \delta \left ( k^2 - x_{34} M_1^2 \right )\, \frac{d^4 k}{(2 \pi)^4}\,
(2 \pi)^4\,\delta^4 \left ( p - q - k \right ), \nonumber 
\\
d\Phi_2^{(12)} &=&2 \pi \delta \left ( k_1^2 -\, m_{\ell}^2 \right )\, \frac{d^4 k_1}{(2 \pi)^4}\,
2 \pi \delta \left ( k_2^2 -\, m_{\ell}^2 \right )\, \frac{d^4 k_2}{(2 \pi)^4}\,
(2 \pi)^4\,\delta^4 \left ( q - k_1 - k_2 \right ), \nonumber
\\
d\Phi_2^{(34)} &=& 2 \pi \delta \left ( k_3^2 \right )\, \frac{d^4 k_3}{(2 \pi)^4}\,
2 \pi \delta \left ( k_4^2 -\, m_{\ell'}^2 \right )\, \frac{d^4 k_4}{(2 \pi)^4}\,
(2 \pi)^4\,\delta^4 \left ( k - k_3 - k_4 \right ) \nonumber
\end{eqnarray} 
It is suitable to choose $x_{12}$, $x_{34}$, $y_{12}$, $y_{34}$, and
$\varphi$ as independent integration variables
when calculating the four-body phase space. Then
$$
\Phi_2^{(qk)} =\,\frac{1}{2^3\, \pi}\,\lambda^{1/2} \left ( 1,\,  x_{12},\, x_{34}\right );
$$
and
$$
d\Phi_2^{(12)} =\,\frac{1}{2^4\, \pi}\,\sqrt{1\, -\,\frac{4 {\hat m}_\ell^2}{x_{12}}}\, d y_{12};\quad
d\Phi_2^{(34)} =\,\frac{1}{2^5\, \pi^2}\,\left ( 1\, -\,\frac{{\hat m}_{\ell'}^2}{x_{34}}\right )  d y_{34}\, d\varphi.
$$
This gives 
\begin{eqnarray}
\label{dPhi1234}
d\Phi_4^{(1234)} &=&                                                                
\frac{M_1^4}{2^{14}\, \pi^6}\,
\lambda^{1/2} \left ( 1,\,  x_{12},\, x_{34}\right )\,\sqrt{1\, -\,\frac{4 {\hat m}_\ell^2}{x_{12}}}\,\,\left ( 1\, -\,\frac{{\hat m}_{\ell'}^2}{x_{34}}\right )
d x_{12}\, d x_{34}\, d y_{12}\, d y_{34}\, d \varphi, 
\end{eqnarray} 
where ${\hat m}_\ell = m_\ell /M_1$ and  ${\hat m}_{\ell'} = m_{\ell'}/M_1$.

In the decay 
$
B^-(p) \,\to\, \ell^+(k_1)\,\ell^-(k_2)\, {\bar \nu}_{\ell}(k_3)\,\ell^-(k_4) 
$ 
there are two identical leptons $\ell^- (k_2)$ and $\ell^- (k_4)$ in
the final state, so Fermi antisymmetrization of the decay amplitude is
necessary by four--momenta $k_2$ and $k_4$.  We will need an
additional formula to calculate of the branching ratio in this case
for $m_\ell \ne 0$:
\begin{eqnarray}
\label{dPhi1432}
d\Phi_4^{(1432)} &=&                                                                
\frac{M_1^4}{2^{14} \pi^6}\,
\lambda^{1/2} \left ( 1,\,  x_{14},\, x_{23}\right )\,\sqrt{1\, -\,\frac{4 {\hat m}_\ell^2}{x_{12}}}\,\,\left ( 1\, -\,\frac{{\hat m}_\ell^2}{x_{34}}\right )
d x_{14}\, d x_{23}\, d y_{14}\, d y_{23}\, d \tilde\varphi, 
\end{eqnarray} 
where $\tilde\varphi$  is the angle of planes $({\bf
  k}_1,\, {\bf k}_4)$ and $({\bf k}_2,\, {\bf k}_3)$,
measured relative to plane 
$({\bf k}_1,\, {\bf k}_4)$. The equation ~(\ref{dPhi1432}) may be obtained
in a fully analogous way to (\ref{dPhi1234}). The
$\cos\tilde\varphi$ can be found by exchanging indices in equations (\ref{cosvarphi-3lnu}) and
(\ref{cosvarphi-equation}) as $2 \leftrightarrow 4$. Also in order to
perform numerical integration it is necessary to have all the
definitions of $x_{ij}$ using the set of variables $x_{12}$, $x_{34}$,
$y_{12}$, $y_{34}$, and $\varphi$.
From (\ref{monentum-equation}), (\ref{costheta-equation}), and
(\ref{cosvarphi-equation}), assuming zero lepton masses, we have:
\begin{eqnarray}
\label{xij-vsakie}
x_{13} &=& \frac{1}{4}\,
\Big (
-2\,\sqrt{x_{12}\, x_{34}\, (1 - y^2_{12})\, (1 - y^2_{34})}\,\cos\varphi\, + (1 - x_{12} - x_{34})\, y_{12}\, y_{34}\, -
\nonumber \\
&-& \lambda^{1/2} \left ( 1,\,  x_{12},\, x_{34}\right )\, (y_{12} + y_{34})\, +\, 1\, -\, x_{12}\, -\, x_{34}
\Big ); \nonumber 
\\
x_{14} &=& \frac{1}{4}\,
\Big (
2\,\sqrt{x_{12}\, x_{34}\, (1 - y^2_{12})\, (1 - y^2_{34})}\,\cos\varphi\, - (1 - x_{12} - x_{34})\, y_{12}\, y_{34}\, -
\nonumber \\
&-& \lambda^{1/2} \left ( 1,\,  x_{12},\, x_{34}\right )\, (y_{12} - y_{34})\, +\, 1\, -\, x_{12}\, -\, x_{34}
\Big ); 
\\
x_{23} &=& \frac{1}{4}\,
\Big (
2\,\sqrt{x_{12}\, x_{34}\, (1 - y^2_{12})\, (1 - y^2_{34})}\,\cos\varphi\, - (1 - x_{12} - x_{34})\, y_{12}\, y_{34}\, +
\nonumber \\
&+& \lambda^{1/2} \left ( 1,\,  x_{12},\, x_{34}\right )\, (y_{12} - y_{34})\, +\, 1\, -\, x_{12}\, -\, x_{34}
\Big ); \nonumber
\\
x_{24} &=& \frac{1}{4}\,
\Big (
-2\,\sqrt{x_{12}\, x_{34}\, (1 - y^2_{12})\, (1 - y^2_{34})}\,\cos\varphi\, + (1 - x_{12} - x_{34})\, y_{12}\, y_{34}\, +
\nonumber \\
&+& \lambda^{1/2} \left ( 1,\,  x_{12},\, x_{34}\right )\, (y_{12} + y_{34})\, +\, 1\, -\, x_{12}\, -\, x_{34}
\Big ); \nonumber 
\end{eqnarray}
This paper use notations almost  identical
to the notations of Ref. \cite{Barker:2002ib}, except for in case of the 
$y_{ij}$, which here have the opposite sign compared to
Ref. \cite{Barker:2002ib}.


\end{document}